\documentclass[sigconf,screen,nonacm,pbalance]{acmart}
\author{
  Angela Cui*,
  Ferran Hermida-Rivera*,
  Jack Toubes*, \\[0.5ex]
  Raghav Gupta, 
  Jim Fang,  
  Chengyi Lux Zhang, 
  Ella Schwarz, 
  Junha Kim, \\[0.5ex]
  Yakun Sophia Shao, 
  Borivoje Nikolić, 
  Christopher W. Fletcher,
  Sagar Karandikar
}

\affiliation{%
  \vspace{1ex}
  \institution{
    University of California, Berkeley \\
    \vspace{1ex}
  }
  \country{} 
}

\email{{angelacui, ferran, jtoubes, raghavgupta, jimfang, luxz, schwarzem, junhak}@berkeley.edu}
\email{{ysshao, bora, cwfletcher, sagark}@eecs.berkeley.edu}

\thanks{\textsuperscript{*}Equal Contribution}

\copyrightyear{}
\acmYear{}
\acmDOI{}
\acmConference[]{}{}{}

\acmISBN{}

\usepackage{amsmath,amssymb,amsfonts}
\usepackage{algorithmic}
\usepackage{graphicx}
\usepackage{textcomp}
\usepackage{xcolor}
\usepackage{circuitikz}
\usepackage{pgfplots, pgfplotstable}
\usepackage{tabularx}
\usepackage{float}
\usepackage[font=small,labelfont=bf]{subcaption}
\usepackage{array}
\usepackage{layouts}
\usepackage{makecell}
\usepackage{placeins}
\usepackage{comment}
\usepackage{xurl}
\usepackage{hyperref}
\usepackage{tabularx,tabularkv,lipsum}
\usepackage{svg}
\usepackage{minted, listings}
\usepackage[font=small,labelfont=bf]{caption}
\usepackage[normalem]{ulem}
\useunder{\uline}{\ul}{}
\usepackage[normalem]{ulem}
\usepackage[autostyle, english=american]{csquotes}
\MakeOuterQuote{"}

\def\BibTeX{{\rm B\kern-.05em{\sc i\kern-.025em b}\kern-.08em
    T\kern-.1667em\lower.7ex\hbox{E}\kern-.125emX}}

\makeatletter
\newcommand{\thickhline}{%
    \noalign {\ifnum 0=`}\fi \hrule height 1pt
    \futurelet \reserved@a \@xhline
}

\makeatletter
\newcommand\blfootnote[1]{%
  \begingroup
  \renewcommand\thefootnote{}
  \protected@xdef\@thefnmark{}
  \@footnotetext{#1}
  \endgroup
}

\usepackage[colorinlistoftodos,prependcaption,color=yellow]{todonotes}

\definecolor{ClearSky}{RGB}{173, 237, 255}
\definecolor{Mist}{RGB}{232, 250, 255}
\definecolor{Cereal}{RGB}{255, 234, 150}
\definecolor{Wheat}{RGB}{255, 249, 223}
\definecolor{OceanDepth}{RGB}{26, 31, 161}
\definecolor{Abyss}{RGB}{15, 19, 96}
\definecolor{Spearmint}{RGB}{149, 251, 190}
\definecolor{Lichen}{RGB}{232, 253, 240}
\definecolor{Eraser}{RGB}{255, 195, 254}
\definecolor{RoseOfSharon}{RGB}{255, 233, 255}
\definecolor{HotPink}{RGB}{247, 0, 103}
\definecolor{HotBlue}{RGB}{13, 101, 255}
\definecolor{HotGreen}{RGB}{19, 180, 10}
\definecolor{HotOrange}{RGB}{255, 87, 51}
\definecolor{LightGray}{gray}{0.65}
\definecolor{MediumGray}{gray}{0.45}

\newif\ifshowcomments
\showcommentstrue
\ifshowcomments

\else
  \newcommand{\todo}[1]{}
\fi
\begin{document}

\newcommand{\pyinline}[1]{\lstinline[language=Python, basicstyle=\ttfamily, breaklines=true, breakatwhitespace=false]|#1|}

\title[CHIA: An open-source framework for principled, agentic AI-driven hardware/software co-design research]{CHIA: An open-source framework for principled, agentic\\AI-driven hardware/software co-design research}

\setcopyright{cc}
\setcctype{by}

\begin{abstract}
Agentic artificial intelligence shows great promise for radically improving the pace of innovation in hardware/software co-design research across computer architecture, systems, compilers, and VLSI. Thus far, however, applications of AI in these contexts have generally been demonstrated in isolated settings on small-scale problems, due to the difficulty of designing and deploying complex AI-infused hardware and software development workflows.

This paper introduces CHIA, an open-source hardware/software co-design framework for agile and principled research on the application of AI to co-design. CHIA treats the productive construction and scalable deployment of the co-design flow itself as a first-class objective. In CHIA, agentic AI-driven hardware and software design flows are expressed as \textit{CHIA loops}: directed cyclic graphs whose nodes execute various system-on-chip design tools, microarchitectural simulators, software build systems, AI models, evolutionary coding agents, and more. The \textit{CHIA library} provides node implementations for many popular tools, including Chipyard, gem5, ChampSim, FireSim, Hammer (thus several commercial ASIC CAD tools), Vivado, AlphaEvolve, AdaEvolve, and many others. 

CHIA also provides a broad set of features to conduct principled science around these flows. These include isolation between AI models and hardware tools, profiling mechanisms, fault-tolerant execution, and reliability at scale across hundreds of heterogeneous systems (CPUs, FPGAs, GPUs, etc., across public cloud/on-prem.).

To showcase CHIA, we present five CHIA loops as case studies: (1) automatic RTL-to-gem5 simulator alignment, (2) LLM-driven implementation of microarchitectural features in RTL, (3) agentic, IPC-aware critical path optimization, (4) evolutionary architectural discovery, and (5) maintainer-friendly agentic GitHub issue fixing.

Across the case studies, the agents operating within CHIA loops are able to meet the high standards of verification and validation that the loops enforce. For example, the RTL implemented by agents delivers substantial performance improvements while successfully executing all 25+ trillion instructions of the SPEC CPU2006 reference suite within an out-of-order superscalar microprocessor in a RISC-V system-on-chip. These agent-written implementations improve performance while meeting or improving frequency and area constraints in open-source/commercial ASIC PDKs.

These case studies are not an exhaustive demonstration of CHIA's capabilities. Rather, we hope they serve as an instructive set of examples for the community to build on.
\end{abstract}
\maketitle

\section{Introduction}\label{sec:intro}

The design of modern computing systems is a challenging and labor-intensive endeavor. Architects, system designers, compiler engineers, and VLSI specialists must navigate enormous design spaces while reasoning across several abstraction layers, including software, architecture, microarchitecture, and physical implementation. Pushing the limits of co-design across these layers remains an important challenge and opportunity.

Recent advances in agentic artificial intelligence offer an opportunity to accelerate the pace of innovation in both hardware design and at the hardware/software boundary. For example, in November 2025, Gupta et. al.~\cite{anonymous2025archagent, Gupta26} demonstrated the ability to harness AI to automatically generate new state-of-the-art microarchitectural components (cache replacement policies) using the context of established design championships. As another case-in-point, Krishna, et. al.~\cite{VerkorTeam26, theverkorteam26_2} recently demonstrated the generation of a 5-stage RISC-V microprocessor design from scratch using agentic AI. Computer architecture has clearly entered a new era, wherein humans, agents, and existing hardware design tools must all smoothly operate together~\cite{Zhang26, anonymous2025archagent, Blasberg26, Deng26, Wu26, Yu25, Bhandwaldar26, Lu25, Bai25, Davis26, VerkorTeam26, Tsai26, Chen26, Sharma26, Sankaralingam26, Surya94, Guo25}. This is a tall order, considering that constructively composing hardware design tools was a challenge even before the application of AI. Moreover, while the aforementioned studies show interesting achievements from applying AI to hardware design and HW/SW co-design problems, they present early and individual points in the space of potential AI-driven design flows. \emph{How} to design the most effective AI-driven design flows remains a critical open question.

\begin{figure*}[!t]
  \centering
  \includegraphics[width=2\columnwidth]{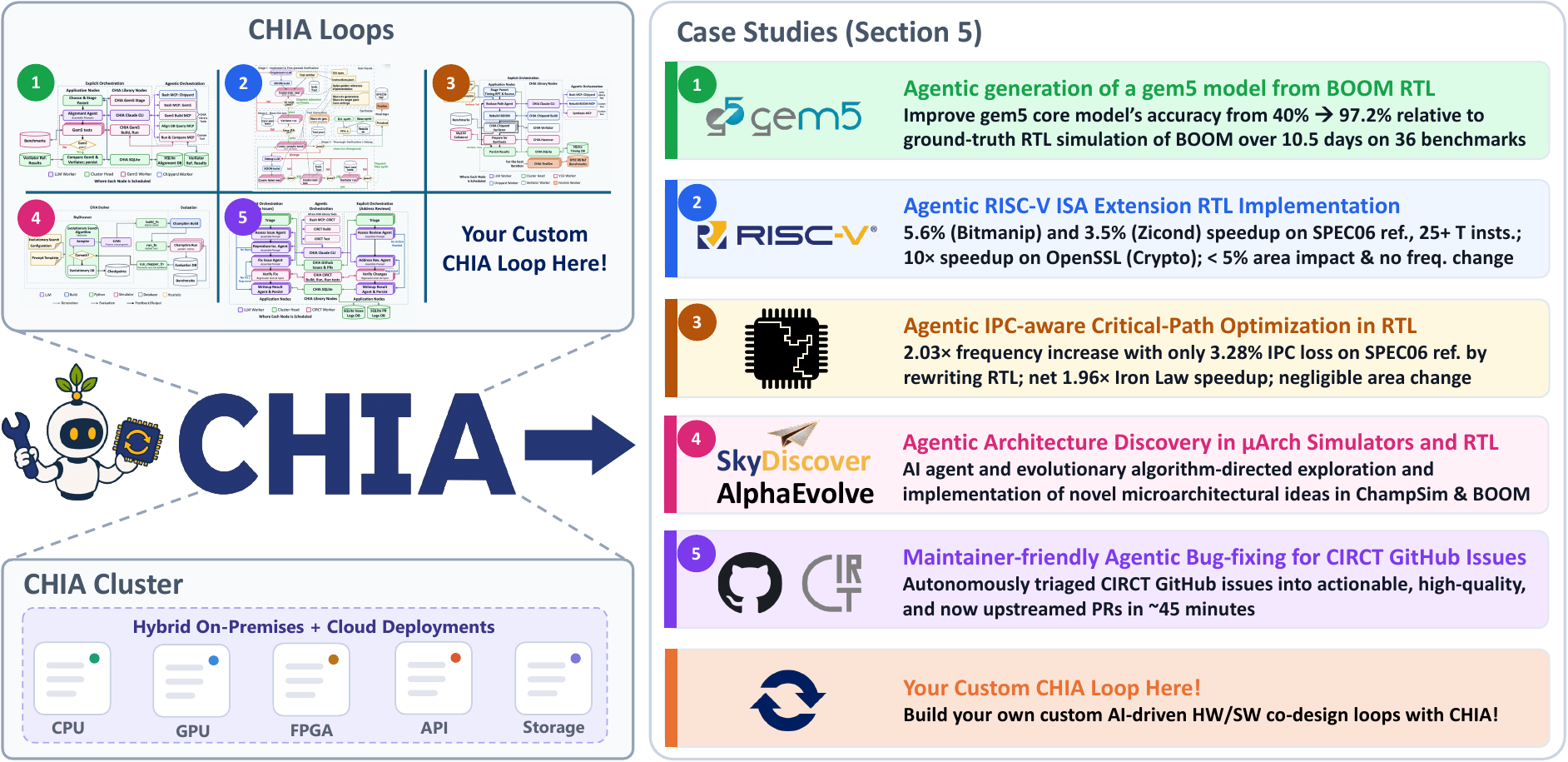}
  \caption{Executive summary of this work.}
\end{figure*}

Naively deploying existing hardware design workflows is not sufficient to tackle this question. The quick-and-dirty approach of writing bespoke scripts that glue together the output of one tool into the input of the next tool (or AI model) breaks easily and does not scale well to large AI-driven design flows on hundreds of heterogeneous machines (CPUs, FPGAs, GPUs, etc.). Shifting from script writing to building a distributed and scalable implementation of each bespoke AI-driven design flow is error-prone and imposes immense engineering cost, even with AI-assistance. AI agents can also orchestrate design flows themselves, sometimes desirably so. However, relying strictly on agents to orchestrate the entire design flow (e.g., letting a coding agent ``run wild'' inside system-on-chip design frameworks~\cite{Amid20, openpiton, Mantovani20, pulp} or microarchitectural simulators~\cite{Binkert11, gem5-new, Gober22, scarab, sniper, zsim}) makes it challenging to enforce verification and validation requirements that are critical to successful hardware design while minimizing human review overhead. Though several design flows have been proposed (e.g., evolutionary coding agents, multi-agent collaborative environments, etc.), they still prescribe a single design pattern, rather than enabling the expression of arbitrary AI-driven hardware design flows. Altogether, there is no good option for productively expressing and reliably deploying new agentic AI-driven hardware/software co-design flows, which hinders the progress of the field.

We built CHIA\footnote{CHIA originally stood for Chipyard Intelligent Agents (or: the Chipyard Intelligence Agency), but was retroactively renamed as we realized its reach was far beyond a single RTL/system-on-chip (SoC) design framework. CHIA now stands for \textbf{C}o-designing \textbf{H}ardware/software with \textbf{I}ntelligent \textbf{A}gents, and supports many common design flows.} to address this problem. CHIA is an open-source, agent-forward hardware/software co-design framework built on the premise that the productive construction and scalable deployment of the co-design flow itself should be a first-class objective. CHIA enables agile and principled research on the application of AI to co-design by making complex AI-infused workflows easy to express, deploy, and study. 
 
In CHIA, the expression of a design flow is called a "CHIA loop"\footnote{CHIA Project Website: \url{https://chialoops.ai}}, which is a directed cyclic graph whose nodes execute various system-on-chip design tools, microarchitectural simulators, software build systems, AI models, evolutionary coding agents, and more. The CHIA library includes node implementations for many widely used tools, including Chipyard, gem5, ChampSim, FireSim, Hammer (and thus several commercial ASIC CAD tools), Vivado, AlphaEvolve, AdaEvolve, and many others, allowing researchers to compose sophisticated flows from reusable components rather than from scratch.

Beyond composition, CHIA provides the features needed to conduct principled science around these flows. These include isolation between AI models and hardware tools, profiling mechanisms, and fault tolerance and reliability at scale across hundreds of heterogeneous systems. Together, these capabilities allow co-design experiments to be run reproducibly and at a scale that makes meaningful evaluation of agent capabilities possible.

To showcase CHIA, we present five example CHIA loops as case studies: (1) automatic, agentic generation of a representative gem5 model for BOOM; (2) LLM-driven implementation of microarchitectural features in RTL in an out-of-order superscalar RISC-V microprocessor, BOOM~\cite{Zhao20}; (3) agentic, IPC-aware critical path optimization; (4) evolutionary architectural discovery flows (à la~\cite{anonymous2025archagent, Gupta26, Blasberg26}); and (5) maintainer-friendly automatic GitHub-issue fixing in the CIRCT compiler project~\cite{eldridge21}. Throughout, we hold the agents to a high standard of verification and validation. For example, when RTL changes are made, we show that RTL implemented by the agents delivers substantial performance improvements while successfully executing the entire 25+ trillion instructions of SPEC06-Ref within a complete out-of-order superscalar microprocessor (BOOM) in a RISC-V system-on-chip (Chipyard)~\cite{Amid20}. These gains are achieved while meeting or improving frequency and area constraints in open-source and commercial ASIC PDKs.

These case studies are not an exhaustive demonstration of what CHIA can do. Rather, we offer them as an instructive set of examples that illustrate the framework's capabilities and that we hope the community will build upon. CHIA is open-source on GitHub \footnote{CHIA GitHub: \url{https://github.com/ucb-bar/chia}} and we are excited to see what the community builds upon it.

\section{CHIA Design Goals}\label{sec:mot}

In this section we highlight the capabilities we want CHIA to provide to its users, to enable them to build productive, agile, AI-driven design workflows.

In an agentic loop targeting a task like the hardware/software co-design of a domain-specific accelerator, agents can significantly reduce the time to perform many relevant tasks, from profiling to RTL implementation. Moreover, agents, unlike humans, are able to perform these tasks at huge scale and can implement tens and even hundreds of designs in parallel. Such massive parallelism is what many LLM-guided search and optimization algorithms, including evolutionary algorithms, leverage. At such a large scale, brute force scripting or bespoke construction of an entire distributed design flow for each new approach becomes infeasible. \textbf{CHIA's graph-based loop abstraction must allow users to build parameterized, massive-scale design pipelines simply.}

At this scale, classic distributed computing challenges such as task scheduling become important. Tasks in hardware design loops occupy a wide range of latencies, from milliseconds to days, and have diverse and strict resource requirements, such as FPGAs for simulation or GPUs for model serving or ASIC CAD tools. Similarly, node failures will impact jobs more frequently at this scale. \textbf{The nodes in CHIA loops should be flexibly and fault-tolerantly scheduled onto a dynamic and heterogeneous substrate.}

As a project progresses, an architect may use many different tools, each with different input and output shapes, and organize them in many different ways. A naive approach that pieces together different tools using bespoke scripts or programs becomes brittle at scale and increasingly difficult to extend as a workflow grows more complex. \textbf{CHIA should be modular, and should provide plug-and-play support for many different co-design tools with various input/output formats.}

Additionally, control flow in a design loop is not always easy to define statically. In an agentic loop, this may manifest as explicit conditional logic guiding the flow. But, as the capabilities of agents improve, their own ability to guide flows can be leveraged to adapt a workflow on the fly. \textbf{CHIA must be able to express programmatic and agentic decision making, and as agents become more capable and potentially more trustworthy, the framework must be able to tune the agent's decision making control.}

AI-driven design brings new opportunities, with radically reduced time from idea to design, but also new challenges. A growing body of work investigates the capabilities of various agentic loops to tackle hardware/software design problems. However, we need principled methods to evaluate these loops. Metrics should be built into agentic platforms for trustworthy and simple data collection. In addition to introspection to improve the loop itself, data collection is also invaluable for evaluation of the design generated by the flow. \textbf{CHIA needs data collection and profiling of both results and of the loop itself built in as first-class citizens.}

Together, these requirements enable us to perform agile and productive experimentation on agentic loops targeting co-design.

\section{The CHIA Workflow Abstraction}\label{sec:arch}
A CHIA project is called a workflow, and is composed of a CHIA cluster and a CHIA loop. \textit{Clusters} are physical machines and software environments where work is executed, and \textit{loops} are programs which \textit{orchestrate} a pipeline of hardware and software design tasks. In this section, we illustrate the CHIA design abstractions which constitute loops and clusters, with the help of a running example.

\subsection{Running Example}\label{sec:work-eg}

\begin{figure*}[!ht]
\centering
\usetikzlibrary{arrows.meta,shapes.symbols,calc}
\resizebox{0.9\textwidth}{!}{%
\begin{circuitikz}
\tikzstyle{every node}=[font=\fontsize{14.2pt}{18.5pt}\bfseries\selectfont]
\draw [ color={rgb,255:red,90; green,90; blue,90}, draw opacity=1, fill={rgb,255:red,250; green,250; blue,250}, fill opacity=1, line width=1.0pt, dashed, rounded corners = 24] (-2,-0.75) rectangle (45,12.4);
\draw [ color={rgb,255:red,90; green,90; blue,90}, draw opacity=1, fill={rgb,255:red,250; green,250; blue,250}, fill opacity=1, line width=1.0pt, dashed, rounded corners = 24] (-2,23.25) rectangle (45,12.75);
\node[anchor=center, align=center, rotate=90, fill={rgb,255:red,250; green,250; blue,250}, fill opacity=1, inner xsep=0.080cm, inner ysep=0.085cm] at (-1.4,18) {\fontsize{19.9pt}{25.9pt}\selectfont \textbf{CHIA Loop}};
\draw [ color={rgb,255:red,74; green,142; blue,79}, draw opacity=1 , fill={rgb,255:red,244; green,244; blue,245}, fill opacity=1, line width=1.75pt , rounded corners = 21.6] (-0.75,12) rectangle (44.25,5.25);
\node[anchor=north, align=center, fill={rgb,255:red,244; green,244; blue,245}, fill opacity=1, text opacity=1, inner xsep=0.080cm, inner ysep=0.085cm, rounded corners=0.020cm] at (21.75,11.85) {\fontsize{19.9pt}{25.9pt}\selectfont \textbf{Logical Workers}};
\draw [ color={rgb,255:red,22; green,163; blue,74}, draw opacity=1, fill={rgb,255:red,220; green,252; blue,231}, fill opacity=1, line width=1.75pt , rounded corners = 6.0] (0,9.75) rectangle (7.5,6);
\node[anchor=center, align=center, inner xsep=0.080cm, inner ysep=0.085cm, rounded corners=0.020cm] at (3.75,7.875) {\fontsize{15.9pt}{20.7pt}\selectfont LLM \\ image: chia-claude-code:latest \\ resource: claude\_creds: 1 \\ num\_workers: 1};
\draw [ color={rgb,255:red,22; green,163; blue,74}, draw opacity=1, fill={rgb,255:red,220; green,252; blue,231}, fill opacity=1, line width=1.75pt , rounded corners = 6.0] (9,9.75) rectangle (16.5,6);
\node[anchor=center, align=center, inner xsep=0.080cm, inner ysep=0.085cm, rounded corners=0.020cm] at (12.75,7.875) {\fontsize{15.9pt}{20.7pt}\selectfont Chisel\_Builder \\ image: chia-chisel-build:latest \\ resource: chipyard: 1, firtool: 1 \\ num\_workers: 1};
\draw [ color={rgb,255:red,22; green,163; blue,74}, draw opacity=1, fill={rgb,255:red,220; green,252; blue,231}, fill opacity=1, line width=1.75pt , rounded corners = 6.0] (36,9.75) rectangle (43.5,6);
\node[anchor=center, align=center, inner xsep=0.080cm, inner ysep=0.085cm, rounded corners=0.020cm] at (39.75,7.875) {\fontsize{15.9pt}{20.7pt}\selectfont Verilator\_Runner \\ image: chia-verilator-run:latest \\ resource: verilator: 1 \\ num\_workers: 3};
\draw [ color={rgb,255:red,22; green,163; blue,74}, draw opacity=1, fill={rgb,255:red,220; green,252; blue,231}, fill opacity=1, line width=1.75pt , rounded corners = 6.0] (27,9.75) rectangle (34.5,6);
\node[anchor=center, align=center, inner xsep=0.080cm, inner ysep=0.085cm, rounded corners=0.020cm] at (30.75,7.875) {\fontsize{15.9pt}{20.7pt}\selectfont Firesim\_Runner \\ image: none \\ resource: FPGA: 1 \\ num\_workers: 1};
\draw [ color={rgb,255:red,22; green,163; blue,74}, draw opacity=1, fill={rgb,255:red,220; green,252; blue,231}, fill opacity=1, line width=1.75pt , rounded corners = 6.0] (18,9.75) rectangle (25.5,6);
\node[anchor=center, align=center, inner xsep=0.080cm, inner ysep=0.085cm, rounded corners=0.020cm] at (21.75,7.875) {\fontsize{15.9pt}{20.7pt}\selectfont Bitstream\_Builder \\ image: chia-vivado:latest \\ resource: Vivado: 1 \\ num\_workers: 1};
\draw [ color={rgb,255:red,167; green,145; blue,83}, draw opacity=1 , fill={rgb,255:red,244; green,244; blue,245}, fill opacity=1, line width=1.75pt , rounded corners = 21.6] (1.25,4.5) rectangle (8.25,0);
\node[anchor=south, align=center, inner xsep=0.080cm, inner ysep=0.085cm, rounded corners=0.020cm] at (4.75,0.15) {\fontsize{19.9pt}{25.9pt}\selectfont \textbf{Cloud Machines}};
\draw [ color={rgb,255:red,167; green,145; blue,83}, draw opacity=1 , fill={rgb,255:red,244; green,244; blue,245}, fill opacity=1, line width=1.75pt , rounded corners = 21.6] (9.75,4.5) rectangle (29.75,0);
\node[anchor=south, align=center, inner xsep=0.080cm, inner ysep=0.085cm, rounded corners=0.020cm] at (19.75,0.15) {\fontsize{19.9pt}{25.9pt}\selectfont \textbf{Local Physical Machines}};
\draw [ color={rgb,255:red,100; green,116; blue,139}, draw opacity=1, fill={rgb,255:red,248; green,250; blue,252}, fill opacity=1, line width=1.75pt , rounded corners = 6.0] (10.75,3.75) rectangle (14.5,1.5);
\node[anchor=center, align=center, inner xsep=0.080cm, inner ysep=0.085cm, rounded corners=0.020cm] at (12.625,2.625) {\fontsize{18.2pt}{23.7pt}\selectfont CPU1 \\ (CPU node)};
\draw [ color={rgb,255:red,100; green,116; blue,139}, draw opacity=1, fill={rgb,255:red,248; green,250; blue,252}, fill opacity=1, line width=1.75pt , rounded corners = 6.0] (15.5,3.75) rectangle (19.25,1.5);
\node[anchor=center, align=center, inner xsep=0.080cm, inner ysep=0.085cm, rounded corners=0.020cm] at (17.375,2.625) {\fontsize{18.2pt}{23.7pt}\selectfont CPU2 \\ (CPU node)};
\draw [ color={rgb,255:red,233; green,113; blue,50}, draw opacity=1, fill={rgb,255:red,246; green,198; blue,173}, fill opacity=1, line width=1.75pt , rounded corners = 6.0] (20.25,3.75) rectangle (24,1.5);
\node[anchor=center, align=center, inner xsep=0.080cm, inner ysep=0.085cm, rounded corners=0.020cm] at (22.125,2.625) {\fontsize{23.9pt}{31.1pt}\selectfont FPGA1};
\draw [ color={rgb,255:red,100; green,116; blue,139}, draw opacity=1, fill={rgb,255:red,248; green,250; blue,252}, fill opacity=1, line width=1.75pt , rounded corners = 6.0] (25,3.75) rectangle (28.75,1.5);
\node[anchor=center, align=center, inner xsep=0.080cm, inner ysep=0.085cm, rounded corners=0.020cm] at (26.875,2.625) {\fontsize{18.2pt}{23.7pt}\selectfont CPU3 \\ (CPU node)};
\draw [ color={rgb,255:red,100; green,116; blue,139}, draw opacity=1, fill={rgb,255:red,248; green,250; blue,252}, fill opacity=1, line width=1.75pt , rounded corners = 6.0] (0,16.5) rectangle (3.75,13.5);
\node[anchor=center, align=center, inner xsep=0.080cm, inner ysep=0.085cm, rounded corners=0.020cm] at (1.875,15) {\fontsize{19.9pt}{25.9pt}\selectfont 1. HW Spec \\ \fontsize{19.9pt}{25.9pt}\selectfont \& Prompt};
\draw [line width=2pt, -{Stealth[scale=1.5]}, ] (3.75,15) -- (6,15);
\draw [ color={rgb,255:red,219; green,39; blue,119}, draw opacity=1, fill={rgb,255:red,255; green,241; blue,247}, fill opacity=1, line width=1.75pt , rounded corners = 6.0] (6.75,21) rectangle (11.25,18);
\node[anchor=center, align=center, inner xsep=0.080cm, inner ysep=0.085cm, rounded corners=0.020cm] at (9,19.5) {\fontsize{19.9pt}{25.9pt}\selectfont 2. Chisel \\ \fontsize{19.9pt}{25.9pt}\selectfont Read/Write};
\draw [line width=2pt, -{Stealth[scale=1.5]}, ] (9,16.5) -- (9,18);
\draw [line width=2pt, -{Stealth[scale=1.5]}, ] (34.5,15) -- (36.75,15);
\draw [line width=2pt, short] (34.5,19.5) -- (35.25,19.5);
\draw [line width=2pt, short] (35.25,19.5) -- (35.25,15);
\draw [line width=2pt, short] (41.25,16.5) .. controls (41.25,18.75) and (41.25,18.75) .. (41.25,21.75);
\draw [line width=2pt, short] (41.25,21.75) -- (1.5,21.75);
\draw [line width=2pt, -{Stealth[scale=1.5]}, ] (1.5,21.75) -- (1.5,16.5);
\draw [ color={rgb,255:red,37; green,99; blue,235}, draw opacity=1, fill={rgb,255:red,239; green,246; blue,255}, fill opacity=1, line width=1.75pt , rounded corners = 6.0] (6,16.5) rectangle (12,13.5);
\node[anchor=center, align=center, inner xsep=0.080cm, inner ysep=0.085cm, rounded corners=0.020cm] at (9,15) {\fontsize{15.9pt}{20.7pt}\selectfont 2. LLM API Call \\ resources: \\ claude\_creds = 1};
\draw [ color={rgb,255:red,37; green,99; blue,235}, draw opacity=1, fill={rgb,255:red,239; green,246; blue,255}, fill opacity=1, line width=1.75pt , rounded corners = 6.0] (14.25,16.5) rectangle (20.25,13.5);
\node[anchor=center, align=center, inner xsep=0.080cm, inner ysep=0.085cm, rounded corners=0.020cm] at (17.25,15) {\fontsize{15.9pt}{20.7pt}\selectfont Build Chisel \\ resources: \\ firtool = 1};
\draw [ color={rgb,255:red,37; green,99; blue,235}, draw opacity=1, fill={rgb,255:red,239; green,246; blue,255}, fill opacity=1, line width=1.75pt , rounded corners = 6.0] (28.5,16.5) rectangle (34.5,13.5);
\node[anchor=center, align=center, inner xsep=0.080cm, inner ysep=0.085cm, rounded corners=0.020cm] at (31.5,15) {\fontsize{15.9pt}{20.7pt}\selectfont 3. Verilator Sim \\ resources: \\ verilator = 1};
\draw [ color={rgb,255:red,37; green,99; blue,235}, draw opacity=1, fill={rgb,255:red,239; green,246; blue,255}, fill opacity=1, line width=1.75pt , rounded corners = 6.0] (36.75,16.5) rectangle (42.75,13.5);
\node[anchor=center, align=center, inner xsep=0.080cm, inner ysep=0.085cm, rounded corners=0.020cm] at (39.75,15) {\fontsize{15.9pt}{20.7pt}\selectfont 4. Evaluate Perf. \\ \fontsize{15.9pt}{20.7pt}\selectfont \& Functionality \\ resources: none};
\draw [ color={rgb,255:red,37; green,99; blue,235}, draw opacity=1, fill={rgb,255:red,239; green,246; blue,255}, fill opacity=1, line width=1.75pt , rounded corners = 6.0] (14.25,21) rectangle (20.25,18);
\node[anchor=center, align=center, inner xsep=0.080cm, inner ysep=0.085cm, rounded corners=0.020cm] at (17.25,19.5) {\fontsize{15.9pt}{20.7pt}\selectfont Build Bitstream \\ resources: \\ Vivado = 1};
\draw [ color={rgb,255:red,37; green,99; blue,235}, draw opacity=1, fill={rgb,255:red,239; green,246; blue,255}, fill opacity=1, line width=1.75pt , rounded corners = 6.0] (28.5,21) rectangle (34.5,18);
\node[anchor=center, align=center, inner xsep=0.080cm, inner ysep=0.085cm, rounded corners=0.020cm] at (31.5,19.5) {\fontsize{15.9pt}{20.7pt}\selectfont 3. FireSim Sim \\ resources: \\ FPGA = 1};
\draw [line width=2pt, -{Stealth[scale=1.5]}, dashed] (9,13.5) -- (3.75,9.75);
\draw [line width=2pt, -{Stealth[scale=1.5]}, dashed] (17.25,13.5) -- (12.75,9.75);
\draw [line width=2pt, -{Stealth[scale=1.5]}, dashed] (31.5,13.5) -- (39.75,9.75);
\draw [line width=2pt, dashed] (27.75,10.5) -- (30.75,10.5);
\draw [line width=2pt, -{Stealth[scale=1.5]}, dashed] (30.75,10.5) -- (30.75,9.75);
\draw [line width=2pt, dashed] (24,17.25) -- (24,18);
\draw [ color={rgb,255:red,100; green,116; blue,139}, draw opacity=1, fill={rgb,255:red,248; green,250; blue,252}, fill opacity=1, line width=1.75pt , rounded corners = 6.0] (22.5,18.75) rectangle (26.25,15.75);
\node[anchor=center, align=center, inner xsep=0.080cm, inner ysep=0.085cm, rounded corners=0.020cm] at (24.375,17.25) {\fontsize{19.9pt}{25.9pt}\selectfont 1. Tests \& \\ \fontsize{19.9pt}{25.9pt}\selectfont Ref. Impl.};
\draw [line width=2pt,  -{Stealth[scale=1.5]}, ] (12,15) -- (14.25,15);
\draw [line width=2pt, -{Stealth[scale=1.5]}, ] (20.25,15) -- (28.5,15);
\draw [line width=2pt, -{Stealth[scale=1.5]}, ] (20.25,19.5) -- (28.5,19.5);
\draw [line width=2pt, -{Stealth[scale=1.5]}, ] (17.25,16.5) -- (17.25,18);
\draw [line width=2pt, short] (26.25,17.25) -- (31.5,17.25);
\draw [line width=2pt, -{Stealth[scale=1.5]}, ] (31.5,17.25) -- (31.5,16.5);
\draw [line width=2pt, -{Stealth[scale=1.5]}, ] (31.5,17.25) -- (31.5,18);
\draw [line width=2pt, dashed] (28.5,18.75) -- (27.75,18.75);
\draw [line width=2pt, dashed] (27.75,18.75) -- (27.75,10.5);
\draw [line width=2pt, dashed] (20.25,18.75) -- (21,18.75);
\draw [line width=2pt, dashed] (21,18.75) -- (21.75,18.75);
\draw [line width=2pt, dashed] (21.75,18.75) -- (21.75,12);
\draw [line width=2pt, -{Stealth[scale=1.5]}, dashed] (21.75,11.25) -- (21.75,9.75);
\draw [ color={rgb,255:red,167; green,145; blue,83}, draw opacity=1 , fill={rgb,255:red,244; green,244; blue,245}, fill opacity=1, line width=1.75pt , rounded corners = 21.6] (31.25,4.5) rectangle (42.25,0);
\node[anchor=south, align=center, inner xsep=0.080cm, inner ysep=0.085cm, rounded corners=0.020cm] at (36.75,0.15) {\fontsize{19.9pt}{25.9pt}\selectfont \textbf{Remote Physical Machines}};
\draw [ color={rgb,255:red,100; green,116; blue,139}, draw opacity=1, fill={rgb,255:red,248; green,250; blue,252}, fill opacity=1, line width=1.75pt , rounded corners = 6.0] (32.25,3.75) rectangle (36,1.5);
\node[anchor=center, align=center, inner xsep=0.080cm, inner ysep=0.085cm, rounded corners=0.020cm] at (34.125,2.625) {\fontsize{18.2pt}{23.7pt}\selectfont AWS1 \\ (CPU node)};
\draw [ color={rgb,255:red,100; green,116; blue,139}, draw opacity=1, fill={rgb,255:red,248; green,250; blue,252}, fill opacity=1, line width=1.75pt , rounded corners = 6.0] (37.5,3.75) rectangle (41.25,1.5);
\node[anchor=center, align=center, inner xsep=0.080cm, inner ysep=0.085cm, rounded corners=0.020cm] at (39.375,2.625) {\fontsize{18.2pt}{23.7pt}\selectfont AWS2 \\ (CPU node)};
\draw [line width=2pt, -{Stealth[scale=1.5]}, ] (3.75,6) -- (12.25,3.75);
\draw [line width=2pt, -{Stealth[scale=1.5]}, ] (12.75,6) -- (13.25,3.75);
\draw [line width=2pt, -{Stealth[scale=1.5]}, ] (21.75,6) -- (17.375,3.75);
\draw [line width=2pt, -{Stealth[scale=1.5]}, ] (30.75,6) -- (22.125,3.75);
\draw [line width=2pt, -{Stealth[scale=1.5]}, ] (39.75,6) -- (39.375,3.75);
\draw [line width=2pt, -{Stealth[scale=1.5]}, ] (39.75,6) -- (34.125,3.75);
\draw [line width=2pt, -{Stealth[scale=1.5]}, ] (39.75,6) -- (26.875,3.75);
\node[cloud, draw={rgb,255:red,90; green,90; blue,90}, text=black, line width=1.75pt, fill={rgb,255:red,225; green,234; blue,247}, fill opacity=1, aspect=1.5, minimum width=4cm, minimum height=2.5cm, align=center] (cloudllm) at (4.75,2.7) {\fontsize{15.9pt}{20.7pt}\selectfont LLM\\Provider};
\draw [line width=2pt, dotted, -{Stealth[scale=1.5]}, ] (3.75,6) -- (cloudllm.north);
\draw [ color={rgb,255:red,90; green,90; blue,90}, draw opacity=1, line width=1.0pt, dashed, rounded corners = 24] (-3,-1.75) rectangle (46.25,24.9);
\node[anchor=north, align=center, fill=white, fill opacity=1, inner xsep=0.080cm, inner ysep=0.085cm] at (21.75,24.75) {\fontsize{19.9pt}{25.9pt}\selectfont \textbf{{CHIA Workflow}}};
\node[anchor=center, align=center, rotate=90, fill={rgb,255:red,250; green,250; blue,250}, fill opacity=1, inner xsep=0.080cm, inner ysep=0.085cm] at (-1.4,6) {\fontsize{19.9pt}{25.9pt}\selectfont \textbf{{CHIA Cluster}}};
\end{circuitikz}
}%
\caption{Simple example CHIA workflow for turning a specification into an RTL description of hardware. The CHIA loop takes a specification for a hardware block and in a loop uses an LLM to design the block, then programatically evaluates its functionality and performance. Numbers correspond to enumerated steps in Section~\ref{sec:work-eg} }\label{fig:running-eg}
\end{figure*}
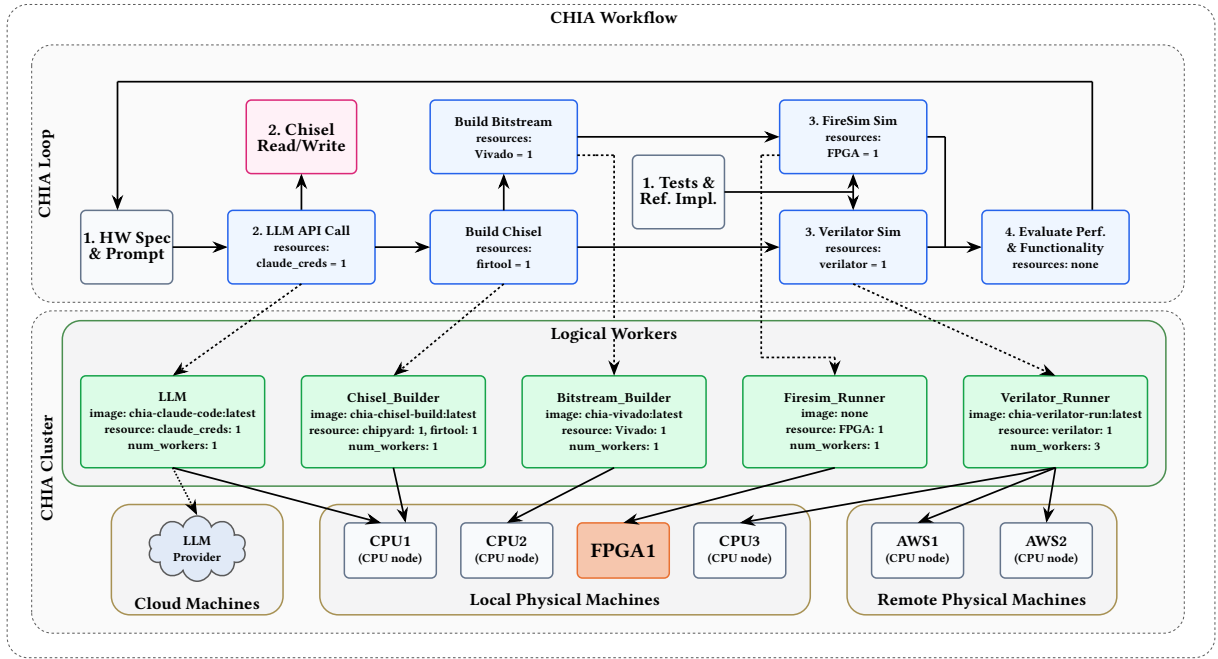

Figure~\ref{fig:running-eg} shows a simple example of one way an architect may use CHIA. Specifically, this workflow uses an agent to write a new hardware module in RTL which achieves higher performance than an existing reference implementation. The steps of the workflow are as follows.

\begin{enumerate}
    \item \textbf{Prepare inputs:} The architect starts with a reference implementation whose performance they want to beat without losing functionality. Next, they write a prompt with the specification of the hardware module, and prepare a suite of unit tests that can be used to verify the functionality of the module.
    \item \textbf{Agentic implementation:} The loop takes the user's prompt and passes it to an agent. The agent attempts to write hardware matching the specification.
    \item \textbf{Simulation:} The loop then compiles a Verilator simulator from the implementation, which is used to run the user's tests. To maximize parallelism, Verilator runs are split across three workers, two of which run on public cloud machines. Some longer benchmarks are run in FPGA-accelerated simulations using FireSim \cite{Karandikar18}.
    \item \textbf{Evaluation:} An evaluator node examines the results of the simulations and determines if the performance threshold is met. If it is not, then the loop restarts. Feedback is provided to the agent, and it continues iterating until the evaluator determines that a stop condition has been reached.
\end{enumerate}

We will use this example in the next few sections to understand CHIA loops and clusters more concretely.

\subsection{Clusters}

Each step in a CHIA workflow may require different physical resources, such as different amounts of memory and different types of accelerators. Similarly, each step may require different software resources, like credentials, dependencies, and isolation, necessitating different environments. CHIA clusters provide the functionality to support these diverse requirements simply and efficiently.

\subsubsection{Physical Machines and Logical Workers}

A cluster starts with a collection of physical machines onto which users plan to distribute the tasks in their CHIA loop. In Figure~\ref{fig:running-eg}, these machines are numbered CPU1 through CPU3, FPGA1, and AWS1 and AWS2. These machines provide heterogeneous physical resources, including different amounts of memory, different processors, and in this example, one machine has an FPGA while the others do not.

On top of these physical machines, users specify a set of \textit{logical} workers. A logical worker's definition includes a set of \textit{resources} which are virtualized representations of the hardware and software environment which a logical worker provides. Nodes in the CHIA loop specify the types and quantities of the resources that they need to run, and, using Ray~\cite{Moritz18}, the CHIA runtime schedules these resource requests to logical workers in the CHIA cluster. This abstraction of resources allows users to abstract certain machine-specific information from the CHIA loop.

Concretely, the \texttt{Firesim\_Runner} logical worker in our example workflow defines a resource called "FPGA", which the \texttt{FireSim Sim} node requests. Similarly, the \texttt{LLM logical worker} defines a resource called "claude\_code", virtualizing a Claude Code binary and credentials, which the \texttt{LLM API Call} node requests.

\subsubsection{Containerization}

CHIA supports and encourages placing logical workers inside of containers. Containerizing workers provides a number of benefits to a workflow. Most importantly for an AI agent-forward platform, containers provide isolation between different workers. An LLM finding and copying from the example workflow's reference implementation of the hardware module would violate the integrity of an experiment. Even worse, an LLM operating in an insecure environment could access privileged information like proprietary PDKs. Assuming a non-adversarial, but potentially erroneous or unreliable LLM, containerization provides this isolation. We do not account for malicious LLMs in our threat model.

Additionally, containerization streamlines first-time cluster setup. Containers front-load and automate a large portion of the one-time setup cost, allowing clusters to be brought up much faster than they otherwise would. For instance, setting up a fresh Chipyard environment for the \texttt{Chisel Builder} worker in our example workflow would normally take nearly an hour, but CHIA's Chipyard container takes seconds to setup.

Containers also make clusters portable, so CHIA clusters can be deployed on a diverse set of physical machines. For instance, not all machines will have Vivado installed, but the example's \texttt{Bitstream Builder} worker can build a bitstream in a container with Vivado on any machine with sufficient computing resources.

\subsubsection{Public Cloud Machines}

CHIA clusters can be split across on-premises/owned computing resources and remote/rented computing resources from the public cloud. This way users can leverage the compute resources they own, while still scaling their experiments. In our example loop, we leverage AWS to run Verilator on more tests than we otherwise could with just on-premises machines.

\subsection{CHIA Loops}

A CHIA loop is a Python program which orchestrates a sequence of tasks required to achieve some hardware or software design-flow related goal. We represent a CHIA loop as a graph, where nodes are the tasks that need to be performed and edges are the data and control flow between tasks. The loop for our running example is described by the graph in Figure~\ref{fig:running-eg}.

\subsubsection{Nodes}\label{sec:arch:nodes} 

A CHIA node is a Python function that performs a step in a hardware/software design pipeline, often by invoking some underlying hardware or software design tool. Each node in our example flow executes an application and performs some postprocessing. For instance, the bitstream build node runs Vivado and collects build logs and results. 

When a node is defined, it is tagged with the resources that it requires from the worker it executes on. Nodes are scheduled and dispatched when all of their inputs are ready and there is a logical worker with sufficient resources to satisfy the node's request. Once the node can be scheduled, its arguments are serialized and sent to the worker where the node will execute. 

Nodes can return control immediately to their callers with a future which can be used to collect their results later. This makes it easy to parallelize work and can be seen in our example workflow, where we can start building a FireSim bitstream at the same time as kicking off multiple Verilator simulations.

Nodes are the basic unit for retry, caching, bypass, subprocess tracking, and profiling, as described later in Section~\ref{sec:runtime}.

\subsubsection{Edges}

Edges represent data and/or control flow in a CHIA graph.
Depending on whether the caller of a node is an AI agent or a line of code in the CHIA loop's Python orchestration program, an edge can either be agentic or programmatic. Any data/control flow that is explicitly specified in the orchestration program of the loop is termed a \textit{programmatic edge}, such as the edge between the \texttt{Build Chisel} and \texttt{Verilator Sim} nodes in our example. An agent using a tool call to run a node, instead of the node being explicitly initiated by the orchestration program, is called an \textit{agentic edge}. In our example workflow, the edge between the LLM and Chisel implementation nodes is an agentic edge. CHIA provides first-class primitives for executing nodes from both programmatic and agentic edges, and users can define any node to be run both programmatically and agentically, or only one or the other. By supporting these two types of edges, CHIA generalizes across a broad spectrum of agentic design patterns that vary in the amount of control and decision-making entrusted to the AI agent.

\section{Design}\label{sec:des}

In this section we detail the design of the interfaces for creating CHIA clusters and loops, and the features of the CHIA runtime which runs a user-defined CHIA workflow.

\subsection{Cluster Interface}

Listing~\ref{list:cluster-yaml} shows part of a configuration file used in CHIA to specify the constituents of the cluster for our running example. Logical workers are defined, along with their exposed resources, container images, and the physical machines they can run on, under the \mintinline[fontsize=\small]{yaml}{available_node_types} mapping. The set of physical machines in the cluster is the union of all IP addresses and hostnames specified for all of the logical workers' \mintinline[fontsize=\small]{yaml}{compatible_ips} keys. The cluster's head machine is specified under the \mintinline[fontsize=\small]{yaml}{provider} mapping. Any cloud instances are specified under a mapping corresponding to the cloud provider (e.g. \mintinline[fontsize=\small]{yaml}{aws_nodes}). 

Clusters are managed from a command line interface (CLI). For example, the cluster in Listing~\ref{list:cluster-yaml} is launched with the command `\mintinline[fontsize=\small]{bash}{chia up Spec2RTL.yaml}' which first spawns any public cloud node instances specified in the yaml, assigns workers to machines, creates containers, and then sets up the workers to listen for nodes. The cluster can be brought down with `\mintinline[fontsize=\small]{bash}{chia down Spec2RTL.yaml}'. 

Clusters can also be expanded dynamically with the CHIA CLI by adding workers to the cluster configuration yaml, and using `\mintinline[fontsize=\small]{bash}{chia up Spec2RTL.yaml --add}', which brings up any workers specified in the configuration file which are not already running. This command can also be used without changing the configuration to attempt to restart dead workers.

Commands required for worker environment setup can be specified per worker in the \mintinline[fontsize=\small]{yaml}{worker_setup_commands} key. Credentials for logging in to each of the machines are specified under the \mintinline[fontsize=\small]{yaml}{auth} mapping, and can be overwritten per physical machine.

\begin{listing}[t]
\begin{minted}[fontsize=\small]{yaml}
# Spec2RTL.yaml

# AWS instances to spawn
aws_nodes:
  verilator_aws:
    InstanceType: c5.9xlarge
    count: 2
    ImageId: ami-0ec10929233384c7f

provider:
  head_ip: CPU1

# Authentication
auth:
    ssh_user: ${USER}
    ssh_private_key: ~/.ssh/key

# Logical Workers
available_node_types:
  Verilator_Runner:
    docker:
      image: "chia-verilator-run:latest"
    resources: {"verilator": 1}
    min_workers: 3
    max_workers: 3
    compatible_ips: [CPU3                \
                     @verilator_aws:0,   \
                     @verilator_aws:1]
    worker_setup_commands: ["source ~/.bashrc"]
...
\end{minted}
\caption{CHIA cluster yaml configuration fragment. Logical workers are specified under the "available\_node\_types" mapping, and physical machines are specified by the union of all logical workers' "compatible\_ips".}
\label{list:cluster-yaml}
\end{listing}

\subsection{Loop Interface}

Listing~\ref{list:flow-api} shows a fragment of a Python file that defines parts of the CHIA loop for our example workflow. The \mintinline[fontsize=\small]{python}{@ChiaFunction} is the core primitive in the CHIA loop. Functions annotated with \mintinline[fontsize=\small]{python}{@ChiaFunction}, like \mintinline[fontsize=\small]{python}{def verilate} in Listing~\ref{list:flow-api} can be scheduled onto the cluster.

\subsubsection{Programmatic Nodes and Edges}

Nodes designed to be run on the cluster specify their resources in their \mintinline[fontsize=\small, breakafter={@}]{python}{@ChiaFunction} annotation as seen in Listing~\ref{list:flow-api}: \mintinline[fontsize=\small]{python}{@ChiaFunction(resources="...")}. A programmatic edge to a node can be created by calling a \mintinline[fontsize=\small]{python}{@ChiaFunction} with \mintinline[fontsize=\small]{python}{fn_name.chia_remote(args)}. A function marked as a \mintinline[fontsize=\small]{python}{@ChiaFunction} can also be run directly like a normal Python function (without \mintinline[fontsize=\small]{python}{chia_remote()}), and it will run on the same machine in the same process as its caller.

\begin{listing}[t]
\begin{minted}[fontsize=\small]{python}
# Spec2RTL.py

# Define node for agentic control flow
@ChiaFunction
def read_chisel_src() -> str:
    ...

@ChiaFunction
def write_chisel_src(contents): 
    ...

class rw_src_tool(ChiaTool):
    def setup(self):
        self.mcp.add_tool(read_chisel_src)
        self.mcp.add_tool(write_chisel_src)

# Prompt the LLM with tools
chisel_impl_Tool = rw_src_tool(
    task_options={"resources": {"firtool": 1.0}})

llm_future = claude_code_cli.prompt.chia_remote(
    prompt, tools=[chisel_impl_Tool])

...

# Define node for programmatic control flow
@ChiaFunction(resources={"verilator": 1.0})
def verilate(sim_binary, test_binaries) 
    -> List(test_result):
    ...

# Schedule the node explicitly
test_results_future = 
    (verilate.chia_remote(sim_bin, test_bins),
     firesim.chia_remote(bitstream, test_bins))

# Pass futures 
evaluate.chia_remote(test_results_future)
# or collect them explicitly
test_results = get(test_results_future)
\end{minted}
\caption{CHIA loop fragment for our running example.}
\label{list:flow-api}
\end{listing}

Nodes called with \mintinline[fontsize=\small]{python}{node.chia_remote(args)} return control immediately to their callers, enabling asynchronous computing. In our example this makes it easy to build a FireSim bitstream and run Verilator simulations in parallel. As shown at the bottom of Listing~\ref{list:flow-api}, these functions return futures which can be blocked on to collect the eventual result using \mintinline[fontsize=\small]{python}{get()}, or can be passed along other explicit edges directly as arguments to other nodes.

\subsubsection{Agentic Nodes and Edges}

As discussed in Section~\ref{sec:arch}, nodes can also be designed to be run by agents as MCP tools. These nodes must be registered with \mintinline[fontsize=\small]{python}{ChiaTool} objects, in a special function which \mintinline[fontsize=\small]{python}{ChiaTool}s must override called \mintinline[fontsize=\small]{python}{def setup(...)}  as shown in Listing~\ref{list:flow-api}. Nodes are defined as functions and attached to a ChiaTool with \mintinline[fontsize=\small]{python}{ChiaTool.mcp.add_tool()}. The ChiaTool object itself sets up an MCP server which hosts all registered tools. This server runs on a worker chosen according to the \mintinline[fontsize=\small]{python}{task_options} argument in its constructor, including resource requirements, specified when initializing the ChiaTool object. 

The \mintinline[fontsize=\small]{python}{ChiaTool.mcp.add_tool()} function takes a bare Python function as input, and by default it executes on the worker that hosts the tool server. This is why we don't need to specify resources for \mintinline[fontsize=\small]{python}{read/write_chisel_src}, since we specify the resources they require when they run when we construct the \mintinline[fontsize=\small]{python}{rw_src_tool}. To register a \mintinline[fontsize=\small]{python}{@ChiaFunction} as a tool, and have it be scheduled remotely at runtime according to its own specified resources, a user would pass \mintinline[
  fontsize=\small,
  breaklines,
  breakafter=.,
  breakafter=(,
  breaksymbolleft={},
  breaksymbolright={}
]{python}{ChiaTool.mcp.add_tool(fn.chia_remote_blocking)}. In this way, a tool server can be hosted on one worker and its tools can execute on another.

All LLM and agent nodes in CHIA expose a prompt method which accepts a list of \mintinline[fontsize=\small]{python}{ChiaTools} which it provides to the agent. The docstring of the registered function becomes the description passed to the LLM. By providing tools for reading and writing Chisel, the LLM in our example loop can decide when in its reasoning it wants to read from and write to the Chisel code.

\subsection{Runtime Features}
\label{sec:runtime}

CHIA's runtime maps a CHIA loop onto a CHIA cluster and handles runtime features like fault tolerance and profiling transparently to the user. In this subsection we discuss some of the features of CHIA's runtime. 

\subsubsection{Mapping Workers to Physical Machines}
Logical workers are mapped onto to physical machines when the cluster is brought online. Multiple workers can share an underlying physical machine (like CPU1 in Figure~\ref{fig:running-eg}), and a single type of worker can be instantiated onto many different physical machines (like our example's Verilator Runner worker). 

Worker definitions specify what machines they are capable of running on. For containerized workers, this is based on what hardware resources the physical machine provides. For bare workers (no container), this also implies that the physical machine has the necessary software resources. The CHIA runtime offers load balancing globally and on a worker by worker basis when allocating workers to compatible physical machines.

\subsubsection{Node Scheduling}
CHIA relies on Ray's scalable distributed scheduling policy to handle the dynamic and sometimes nondeterministic nature of an agentic CHIA loop\cite{Moritz18}. In particular, nondeterministic agents make it impossible to know what nodes will actually run in any given execution of a CHIA loop. At runtime, nodes are scheduled dynamically when all of their inputs are ready and a logical worker with sufficient resources is available.

\subsubsection{Fault Tolerance}\label{sec:cluster:ft}
CHIA leverages Ray's fault-tolerance capabilities in order to reschedule nodes that fail due to system failures \cite{Moritz18}. When a logical worker or a physical machine goes down, that crash is detected and no more work is scheduled on that worker/machine. All tasks which were actively being run by that worker/machine are automatically queued to be rerun if there is another worker/machine with the requested set of resources for the task. Clusters can be expanded dynamically and reintegrate dead nodes with the CHIA CLI using `\mintinline[fontsize=\small]{bash}{chia up --add}', which allows users to add resources to an already running cluster.

\subsubsection{Subprocess Tracking}

CHIA tracks all processes spawned by nodes. When a node or a
loop is stopped/cancelled, CHIA stops all spawned processes that
were not spawned into a new process group, preventing resources
from leaking. If the example workflow was stopped during a long-running bitstream build, this feature would prevent a stray Vivado process from consuming memory for hours after its results stopped being relevant.

\subsubsection{Profiling}

Data collection and loop analytics are built into CHIA through its profiling mechanism. All functions annotated with \mintinline[fontsize=\small]{python}{@ChiaFunction} are automatically profiled, no matter how they are run, and the profiled results are automatically logged to disk. Results include start time, wall-clock time, worker identification, and additional user-specified information. The profiler also reconstructs the task graph by tracking object identity across different node executions. 

\subsubsection{Caching and Bypassing}

CHIA supports bypassing nodes by injecting data into the CHIA loop to stand in for the node's output. Bypassed nodes can still be scheduled according to their resource requirements, enabling easy testing of the CHIA loop without needing to actually run long-running nodes. CHIA also supports a complementary caching feature, where the results of user-specified nodes are persisted on disk. When bypass and caching are used together, CHIA loops can be restarted quickly after crashing by using the cached results to bypass actual node executions. Additionally, caching and bypass can be used to skip over non-deterministic nodes, such as the LLM API call in our example workflow, in order to reproduce an experiment's results.

\subsection{The CHIA Library}

CHIA seamlessly pieces together a diverse set of platforms to provide feature-rich support for a wide range of workflows. As evidence of this, CHIA provides a library of reusable CHIA nodes and supporting containers. We will be continuously adding new features to the CHIA library and, in open-source spirit, welcome contributions from the broader community. The following list contains some of the nodes we already provide as part of the CHIA library:

\begin{itemize}
    \item \textbf{LLM providers, servers, and agents: }Vertex AI/Gemini Enterprise Agent Platform~\cite{Gertenhaber26}, Antigravity~\cite{Antigravity}, AWS Bedrock~\cite{Bedrock}, Claude Code~\cite{claudecode}, Codex~\cite{Codex}, Ollama~\cite{ollama23}, OpenCode~\cite{anomalyco25}, vLLM~\cite{Kwon23}, FireworksAI~\cite{Fireworks}, Groq~\cite{Groq}, OpenRouter~\cite{OpenRouter}
    \item \textbf{SoC Design: }Chipyard~\cite{Amid20}
    \item \textbf{Hardware Compilation: }CIRCT~\cite{eldridge21}, Scala FIRRTL Compiler~\cite{Li16, Izraelevitz17}
    \item \textbf{Simulators: }Verilator~\cite{SnyderVerilator}, FireSim~\cite{Karandikar18}, Gem5~\cite{Binkert11}, ChampSim~\cite{Gober22}
    \item \textbf{Verification: }RISC-V torture~\cite{Lee12}, Spike co-simulation~\cite{riscv-software-src11}
    \item \textbf{VLSI: }Hammer (and therefore support for various commercial CAD tools~\cite{Liew22})
    \item \textbf{Databases and Cloud Storage:} Postgres~\cite{Stonebraker86}, SQLite~\cite{Gaffney22}, AWS S3~\cite{AWSS3}
    \item \textbf{Repository Management: }GitHub~\cite{Github}
\end{itemize}

\subsection{Implementation Stack}

To provide this flexible and feature-rich environment, we build CHIA on top of a curated set of existing tools and libraries. The Ray distributed computing platform~\cite{Moritz18} serves as the substrate underlying CHIA's loop, cluster, and runtime implementation. Ray was chosen over other orchestration frameworks like LangGraph~\cite{langgraph}, Microsoft Agent Framework~\cite{microsoft25}, and Apache Airflow~\cite{apache15} because Ray has expressive control flow semantics, fine-grained flexible scheduling, distributed execution support, and distributed fault tolerance support. 

Other projects play significant roles in CHIA's implementation. We leverage Docker for containerization of software resources~\cite{Merkel14}. CHIA MCP tools rely on PrefectHQ's fastmcp~\cite{PrefectHQ24}, and our cloud integration relies on AWS' Boto3~\cite{boto14} and GCP's Python Cloud Client Libraries~\cite{Google14}. Our profiling and visualization mechanisms are built on TensorBoard~\cite{abadi16}, Weights and Biases~\cite{wandb17}, and GraphViz~\cite{Gansner00}.

\section{Case Studies}\label{sec:eval}
We demonstrate CHIA's capabilities through a series of case studies that showcase the flexibility and productivity advantages of the framework. These case studies cover a wide variety of workflows that are critical to research and development in computer architecture and systems. These include:
\begin{enumerate}
\item Generating a representative gem5 microarchitectural simulator from RISC-V BOOM RTL.
\item Implementing RISC-V ISA extensions in an OoO core
\item Automated critical path optimization in an OoO core
\item Agentic discovery of microarchitectural structures with evolutionary coding agents
\item Addressing GitHub issues in the CIRCT compiler in a maintainer friendly way
\end{enumerate}

\subsection{Automatically generating a representative gem5 microarchitectural simulator from RISC-V BOOM RTL}\label{sec:eval:align}
High-level microarchitectural simulators like gem5~\cite{Binkert11} are widely-used tools for exploring architectural ideas. Unfortunately, there is no easy way to accurately map a full RTL description of a processor to a microarchitectural simulator. Large teams dedicate many engineer-hours to keeping simulators up-to-date and validating alignment with their microprocessor implementations~\cite{Black98, Surya94, Rotithor13, Caculo25}. For small teams and academia, this maintenance is not feasible, and microarchitectural simulators either fall out of alignment, or do not exist for RTL implementations.

\begin{figure}
    \centering
    \includegraphics[width=1\columnwidth]{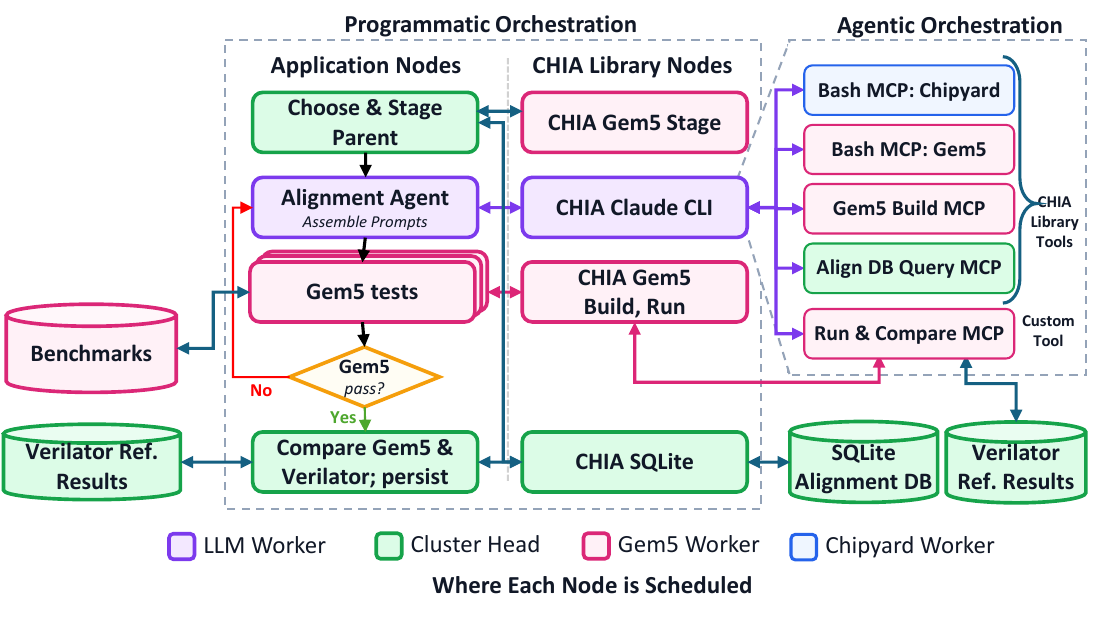}
    \caption{CHIA loop for automatically generating a representative gem5 microarchitectural simulator from RISC-V BOOM RTL.}
    \label{fig:gem5flow}
\end{figure}

We developed a CHIA loop for deriving a gem5-based microarchitectural simulator from an RTL description of a processor, as described in Figure~\ref{fig:gem5flow}. Our pipeline gives an agent (Claude Code with Opus 4.6) access to the source of a gem5 processor core and a gem5 configuration file (where parameters and constants are specified for the core). In each iteration, the agent is asked to modify the architecture and microarchitecture of the gem5 core (not just parameter tuning) to align its performance more closely with a target processor's RTL.

The agent's updated gem5 implementation is executed, and its results are compared to RTL simulation results collected in Verilator. Results are saved into a database, and the parent for the next iteration is chosen based on the most aligned entries in the database. In each iteration, the agent is provided access to this database, including pipeline traces and performance counter results from previous gem5 runs and RTL simulations. We include in our prompt instructions for how to analyze these pipeline traces and performance counters the way a human architect would.

\begin{figure}[t]
    \centering
    \resizebox{\columnwidth}{!}{\input{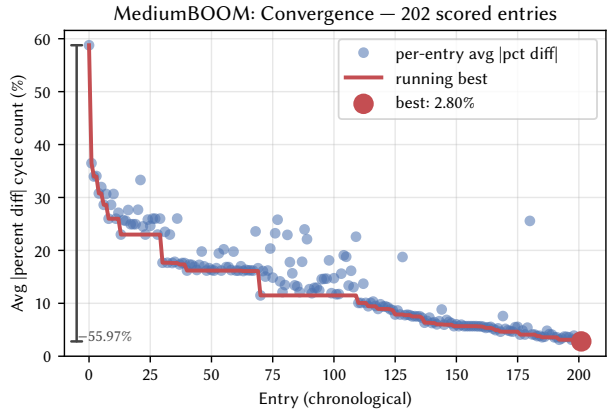}}
    \caption{Average difference in gem5 and Verilator simulation cycle counts for each iteration of the CHIA loop across a suite of small benchmarks. Lower is more aligned. Alignment generally improves as the loop progresses.}
    \label{fig:medboomG5align}
\end{figure}

As shown in Figure~\ref{fig:medboomG5align}, over 202 iterations (about 10.5 days of wall-clock time) our CHIA loop modeled a Chipyard SoC~\cite{Amid20} with a 2-wide MediumBOOM (Berkeley Out of Order Machine) core~\cite{Zhao20, Celio18} in gem5 to within 3\% cycle count error on a training suite of 36 benchmarks taken from the Microbench suite~\cite{Nowatzki15}, each designed to stress different parts of the core. We hypothesized that a large suite of small benchmarks requires a more accurate alignment to reach performance parity than a small suite of large benchmarks, since performance overestimates and underestimates cannot cancel each other out. In reading the LLM's chain of thought, we find that its success arose from comparing pipeline traces and performance counters between the Verilator results and the gem5 results. It was able to use the diverging metrics to correctly identify misaligned structures and align them. The final aligned core has a diff of 1520 lines of code (excluding comments and blank lines) relative to the baseline gem5 RISC-V OoO core.

\begin{figure}[t]
    \centering
    \resizebox{\columnwidth}{!}{\input{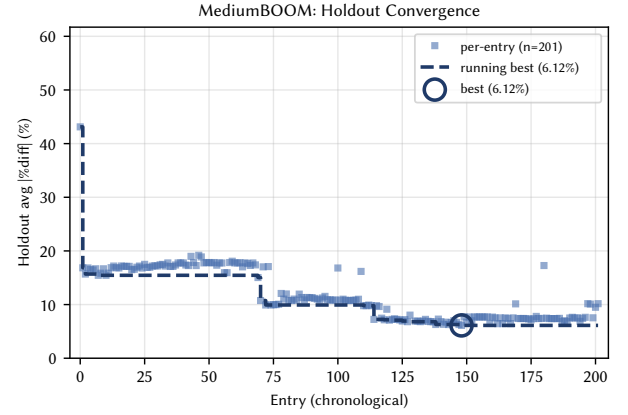}}
    \caption{Average difference in gem5 and Verilator cycle counts for each iteration of the CHIA loop across withheld Embench benchmarks. Lower is more aligned. Alignment generally improves as the loop progresses.}
    \label{fig:medboomG5Withheld}
\end{figure}

We used the Embench IOT 2.0 benchmark suite~\cite{Patterson25} as a withheld and hidden test set to evaluate the quality of our alignment---the agent in the loop does not know Embench is used, cannot see the Embench suite's source, and cannot see its Embench results. In Figure~\ref{fig:medboomG5Withheld}, we show that average misalignment is less than 7\% in our best iteration on the holdout benchmarks, though there are a few outlier benchmarks. 

\begin{figure}[t]
    \centering
    \resizebox{\columnwidth}{!}
    {\includegraphics{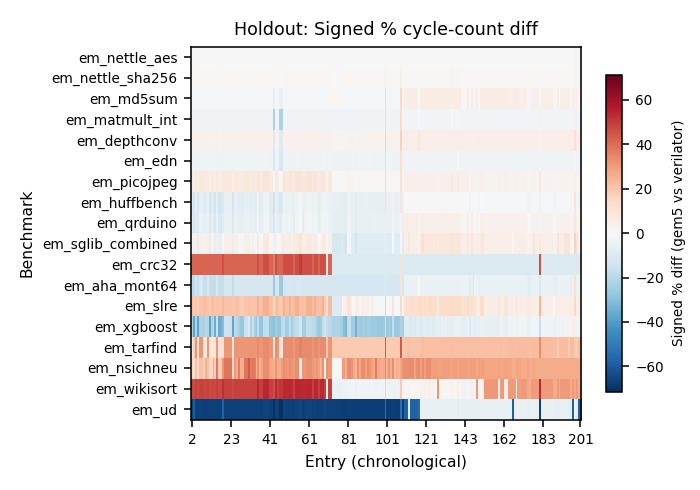}}
    \caption{Per-benchmark and iteration results on the withheld suite. Lighter color is more aligned on a given benchmark in a given iteration. Alignment generally improves on most benchmarks for later iterations.}
    \label{fig:medboomWithheldHeatmap}
\end{figure}

We attribute these outliers primarily to two effects. First, we see instances of overfitting to some of the training set of microbenchmarks. For example, one of the later iterations introduced a parameter \mintinline[fontsize=\small]{c++}{kSuppressThreshold} which suppresses a branch prediction penalty in the gem5 core after a certain number of times the penalty is applied. The agent explicitly stated that the parameter was tuned "with a threshold calibrated to [training benchmark]'s inner loop," to more accurately align that one benchmark. This is the same iteration which introduced a sharp degradation in alignment for the Embench Wikisort benchmark, as can be seen in the bottom right of Figure~\ref{fig:medboomWithheldHeatmap}. The second effect is that our suite of training benchmarks does not capture all potential states of the core. For example, only one of our benchmarks exercises the core with a working set of code larger than the instruction cache. We plan to explore different benchmarking strategies to improve alignment on withheld benchmarks.\footnote{We do not include the SPECint suite for this case study because it is difficult to execute the full SPECint suite in gem5 without SimPoints. In other case studies we are able to do so because we can easily run the full suite on 12 FPGAs in parallel in FireSim.}

On average, each iteration of our loop cost \$11.72 in Claude API credits, as seen in Table~\ref{tab:gem5stats}. We split reporting according to whether the agent could use the "Run \& Compare MCP" tool as described below, as the token usage changes significantly based on whether that tool was used.

\textbf{The Power of CHIA: } We see the power of CHIA around entry 110 in Figure~\ref{fig:medboomG5align}. At this point in the flow, we were not satisfied with the amount of iterations which were regressing overall alignment. To remedy this, we gave the agent the "Run \& Compare MCP" tool, as shown in Figure~\ref{fig:gem5flow}, which gave it the ability to test its gem5 changes on its choice of our training benchmarks. This tool, which was implemented in a single hour, allowed the agent to empirically determine whether its changes were helping and adapt on the fly, which dramatically increased the number of iterations that improved alignment. Not visible in the final results, however, is how easy it was to try many different remedies to address the plateauing results in parallel. In the course of a single day, CHIA's flexibility allowed us to try multiple different strategies for parent selection, prompts for the alignment agent with different selections of information, LLM effort levels, new MCP tools (the "gem5 Build MCP" was introduced shortly before the "Run \& Compare MCP"), and more, before we introduced the "Run \& Compare MCP", which ended up being the most effective.

We provide a detailed walkthrough of the gem5 and BOOM microarchitecture alignment case study on our documentation website.\footnote{\url{http://chialoops.ai/gem5-to-rtl}}


\begin{table}[]
    \centering
    \resizebox{1\columnwidth}{!}{
    \begin{tabular}{lrrrrrr}
    & \multicolumn{2}{c}{All Iterations}
    & \multicolumn{2}{>{\centering\arraybackslash}m{20mm}}{Run \& Compare tool}
    & \multicolumn{2}{>{\centering\arraybackslash}m{20mm}}{No Run \& Compare tool}
    \\\hline
    \multicolumn{1}{c}{Metric} &
    \multicolumn{1}{c}{Avg} &
    \multicolumn{1}{c}{Total} &
    \multicolumn{1}{c}{Avg} &
    \multicolumn{1}{c}{Total} &
    \multicolumn{1}{c}{Avg} &
    \multicolumn{1}{c}{Total} \\\thickhline
        Input Tokens & 151 & 30.3K & 267 & 22.4K & 67 & 7.89K \\
        Cache Creation Tokens & ~300K* & ~60M* & 405K & 34.0M & ~200K* & ~25M* \\
        Cache Read Tokens & ~14M* & ~3B* & 21.0M & 1.76B & ~10M* & ~1B* \\
        Output Tokens & 119K & 23.9M & 189K & 15.9M & 68.4K & 8.0M \\
        Cost (\$) & 11.72 & 2,355.98 & 19.84 & 1,666.48 & 5.89 & 689.50 \\\hline\\
    \end{tabular}
    }
    \caption{Average and total LLM token usage and cost for medium BOOM alignment. Split by LLM access to the Run \& Compare tool which allowed the agent to test its changes. Data with asterisks are estimates, as we did not collect this data until the 113th iteration of the loop. The vast majority of tokens input to the model are consumed as "Cache Creation Tokens" and "Cache Read Tokens", and not as "Input Tokens".}
    \label{tab:gem5stats}
\end{table}

\subsection{Automatically implementing RISC-V ISA extensions in an RTL OOO superscalar core}\label{sec:eval:vext}
Proposing and evaluating microarchitectural improvements is a fundamental part of computer architecture research. Cross-stack approaches often propose new ISA extensions backed by additional microarchitectural features. In either case, implementing these features in RTL and evaluating them on substantial benchmarks in FPGA simulation while collecting ASIC quality-of-result data is the gold standard for evaluation in architecture research. However, despite recent advances in agile hardware design~\cite{Amid20}, realizing this evaluation loop can be challenging for both small chip design teams and pathfinding teams in large companies.

To tackle this challenge, we develop a CHIA loop to implement ISA extensions in RTL for the 4-wide MegaBOOM core in the context of a complete Chipyard SoC. We target the RISC-V Bitmanip~\cite{bitmanip-isa-spec}, Crypto~\cite{crypto-isa-spec}, and Zicond~\cite{zicond-isa-spec} extensions.

First, we demonstrate that in each case, the loop produces a correct implementation, as validated by successfully running substantial benchmarks. 

For Bitmanip and Zicond, this includes all 25.5 trillion instructions of the SPEC06~\cite{SPEC06} reference suite, while demonstrating substantial speedups (5\% and 3.5\% respectively). For Crypto, we demonstrate 10x speedups on OpenSSL benchmarks (200 billion instructions). We also demonstrate that these implementations do not cause critical path regressions in either the SkyWater 130nm process~\cite{SKY130} or a commercial 16nm process, and add only modest silicon area in both cases. 

The CHIA loop takes the specification of a given extension and a formatted list of all of the new instructions and produces a functional and verified implementation with area and timing results. CHIA allows us to make the loop's target highly customizable: a user could implement any extension specification on any core. To do so, the loop only needs to be provided with two pieces of verification collateral: a golden reference functional ISA-simulator implementation (e.g., Spike for RISC-V) for co-simulation and the set of instructions that the core can execute, including the ones to add. The latter collateral will be used by a random test generator to produce a set of random mixes of instructions for functional verification. This collateral is straightforward for the user to implement and provides the basis for verifying the generated RTL.

It is important to note that, within the CHIA loop, verification is as rigorous as in any other open source or academic flow. While the LLM is in charge of implementing and debugging the RTL, the user is still responsible for providing the golden references. CHIA lets us completely isolate the LLM from the verification infrastructure, preventing the agent from modifying the verification infrastructure in any way that would be considered cheating. In fact, the LLM knows nothing about this infrastructure, it only receives the necessary traces and information to debug the RTL.

\begin{figure}
    \centering
    \resizebox{\columnwidth}{!}{\includegraphics[width=\columnwidth]{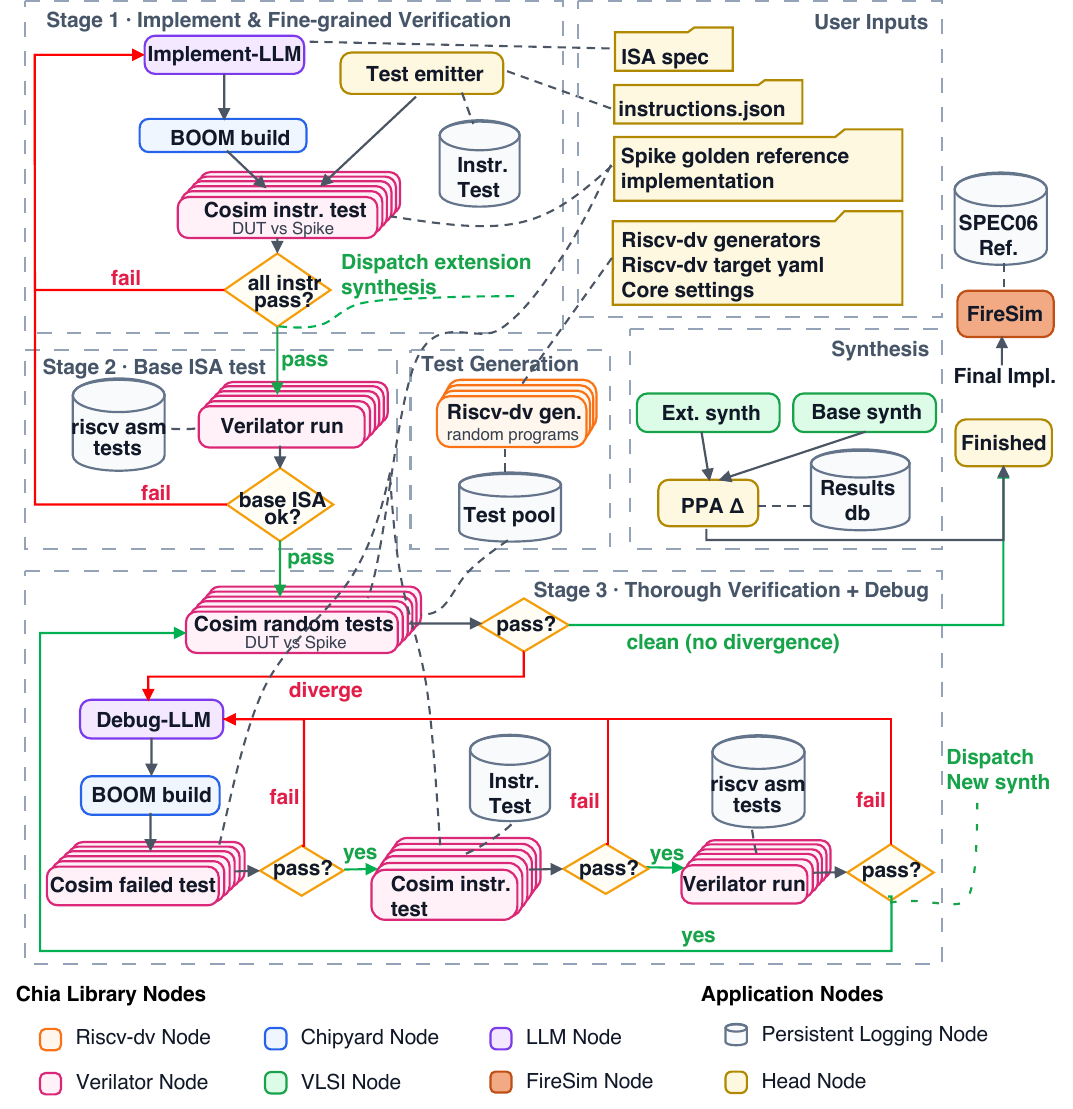}}
    \caption{Architecture of the RISC-V core extension RTL implementation CHIA loop.}
    \label{fig:vextdiagram}
\end{figure}

The loop logic is divided into different stages as depicted in Figure~\ref{fig:vextdiagram}.

\subsubsection{Stage 1}
The first stage consists of a loop with an LLM node in charge of producing the main RTL implementation of a given extension. The LLM uses the user specification to iterate on the RTL. A small set of generated tests that covers the entire repertoire of new instructions is run before moving to the next stage. In parallel, the loop initiates generation of random tests and synthesis of the initial RTL implementation. These random tests consist of long mixes of instructions generated by riscv-dv~\cite{riscvdv} and constitute the basis of Stage 3 verification.

\subsubsection{Stage 2}
In this stage the loop checks for any major regression of the base ISA. It relies on the upstream RISC-V ISA tests~\cite{ASMTest} and in case of an error it goes back to Stage 1 with a trace of the failing test.
\subsubsection{Stage 3}
Only when all instructions execute correctly does the loop move to stage three. It relies on numerous random tests, which were generated in parallel to Stage 1 and 2, that stress the new instructions in combination with the existing ones. In case of divergence it goes back to Stage 1 with a different debugging prompt and the failing test to iterate on an RTL fix. The loop also launches ASIC synthesis of the RTL implementation in this stage.

\begin{figure}[t]
    \centering
    \begin{subfigure}{\columnwidth}\centering
      \includegraphics[width=\columnwidth]{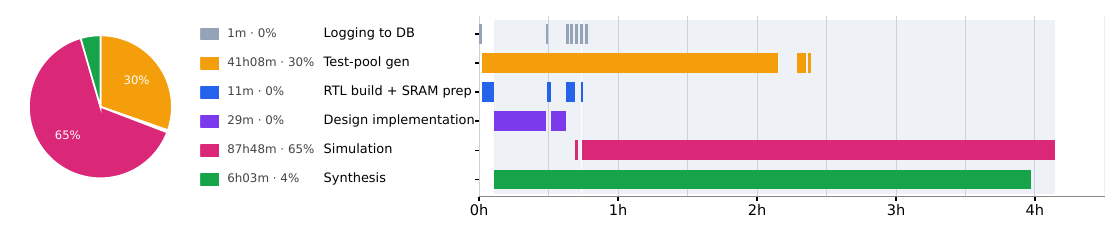}
      \caption{Bitmanip.}\label{fig:tl-bitmanip}
    \end{subfigure}\\[2pt]
    \begin{subfigure}{\columnwidth}\centering
      \includegraphics[width=\columnwidth]{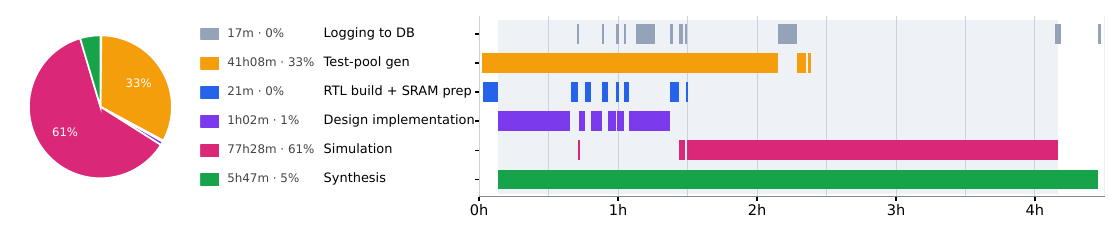}
      \caption{Crypto.}\label{fig:tl-crypto}
    \end{subfigure}\\[2pt]
    \begin{subfigure}{\columnwidth}\centering
      \includegraphics[width=\columnwidth]{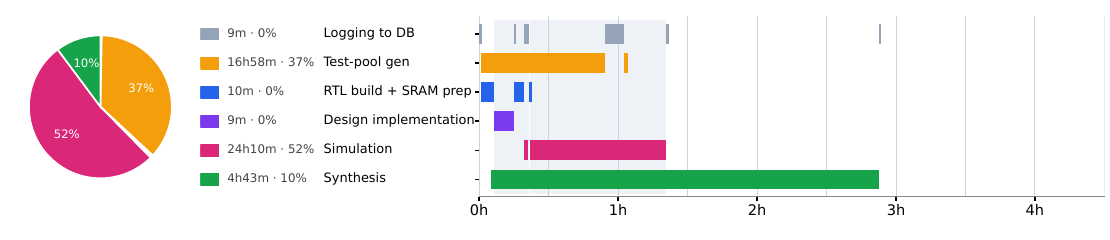}
      \caption{Zicond.}\label{fig:tl-zicond}
    \end{subfigure}
    \caption{Wall-clock execution profile of the agentic RISC-V extension-generation loop. Each panel pairs a Gantt chart of pipeline stages (right) with the share of machine-time per stage (left). Nodes with a total machine time below one minute are not shown (Stage 1's test emitter and PPA collection).}
    \label{fig:ext_timelines}
\end{figure}

With this CHIA loop we automatically implemented and evaluated three new RISC-V extensions in MegaBOOM~\cite{Zhao20,Celio18}; Bitmanip (\mintinline[fontsize=\small]{yaml}{Zba_Zbb_Zbc_Zbs}), Crypto (\mintinline[fontsize=\small]{yaml}{Zbkb_Zbkc_Zbkx_Zknd_Zkne_Zknh}) and Integer Conditional (\mintinline[fontsize=\small]{yaml}{Zicond}). Figure \ref{fig:ext_timelines} represents a summary of the loop evolution constructed using data collected with the CHIA profiler. The timelines show how different nodes were launched and executed while running the loop, and the pie charts on the right show the share of total machine-time per node. Note that some nodes such as simulation and generation run on many machines, while others like synthesis and the LLM run on just one.

Leveraging CHIA's scalability, the loop runs on several heterogeneous machines and enables us to harness as much compute as we need from the cloud, in addition to our local cluster. Bitmanip and Crypto have been implemented on a shared cluster that contained 28 machines (a total of 1024 cores) distributed across both our machines and AWS'. While the LLM implementation phase is lightweight, proper functional verification requires significant compute power to run all the spike co-simulation nodes. In fact, as we can see in Figure ~\ref{fig:ext_timelines}, co-simulation represents between 50\% and 66\% of total machine usage and test generation between a quarter and a third. Zicond was implemented on a smaller cluster of 12 machines and 256 cores, yet its implementation and simulation run considerably faster given that it is the simplest of the three extensions. 

The loop synthesizes both the MegaBOOM~\cite{Zhao20} baseline and the LLM implementations on top of the baseline using the same CHIA synthesis nodes as the loop in Section~\ref{sec:eval:topt}. The area overhead when synthesizing in SkyWater 130nm~\cite{SKY130} is around 2\% for Bitmanip, less than 5\% for Crypto, and is negligible for Zicond. When using a commercial 16nm PDK, the area overhead for Bitmanip is 3\%, for Crypto is 6\%, and for Zicond is 0.8\%. We see no regressions in timing with either PDK.

\begin{figure}[t]
    \centering
    \resizebox{\columnwidth}{!}{\includegraphics[width=\columnwidth]{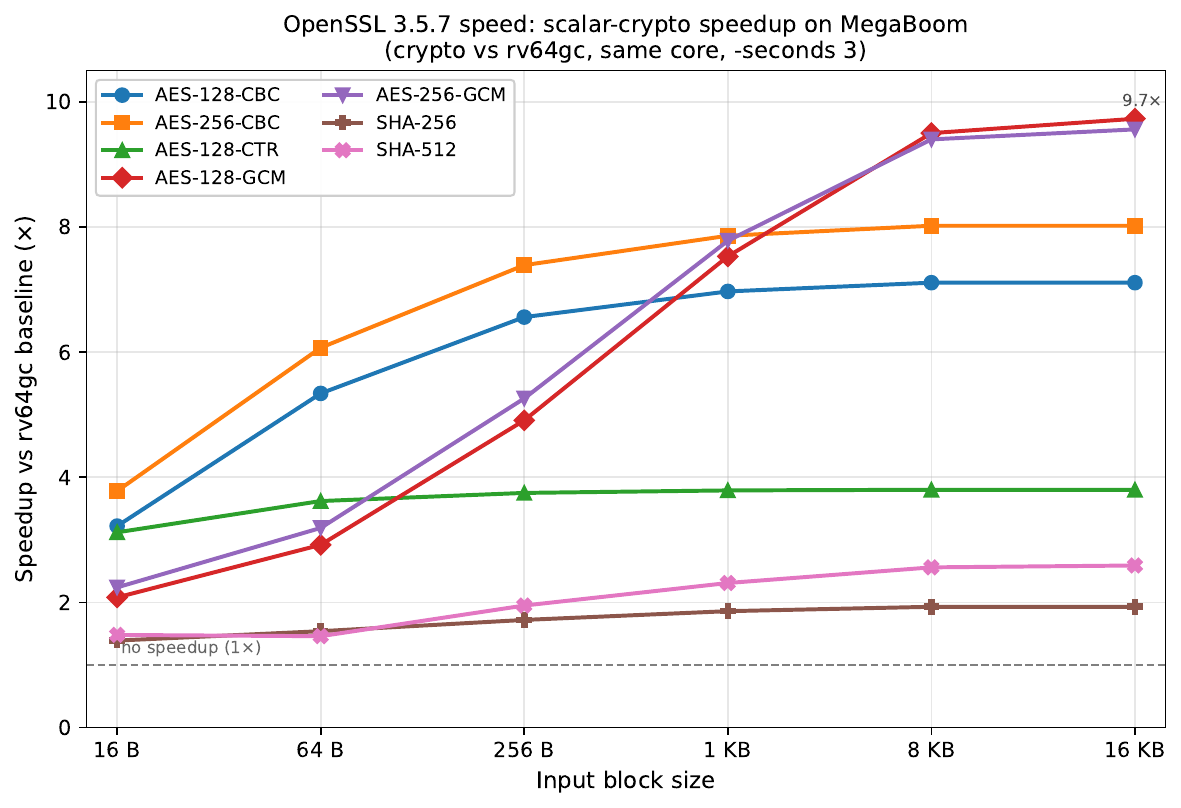}}
    \caption{OpenSSL speedup when executed with the Crypto extension automatically implemented in BOOM using CHIA.}
    \label{fig:openssl}
\end{figure}

  \begin{table}[t]
      \centering
      \resizebox{\columnwidth}{!}{
      \begin{tabular}{lrrrrrr}
      & \multicolumn{3}{c}{Bitmanip (GCC~13.2)}
      & \multicolumn{3}{c}{Zicond (GCC~16.1)} \\\hline
      \multicolumn{1}{c}{Benchmark} &
      \multicolumn{1}{c}{Baseline (s)} &
      \multicolumn{1}{c}{Bitmanip (s)} &
      \multicolumn{1}{c}{Speedup} &
      \multicolumn{1}{c}{Baseline (s)} &
      \multicolumn{1}{c}{Zicond (s)} &
      \multicolumn{1}{c}{Speedup} \\\thickhline
          400.perlbench  & 1501.02 & 1507.84 & 0.995 & 1542.59 & 1461.49 & 1.056 \\
          401.bzip2      & 1593.04 & 1528.20 & 1.042 & 1579.95 & 1570.06 & 1.006 \\
          403.gcc        & 1237.20 & 1209.68 & 1.023 & 1207.76 & 1216.58 & 0.993 \\
          429.mcf        &  997.88 &  997.20 & 1.001 &  974.47 &  974.44 & 1.000 \\
          445.gobmk      & 1317.51 & 1291.24 & 1.020 & 1321.21 & 1316.23 & 1.004 \\
          456.hmmer      & 1972.58 & 1190.03 & 1.658 & 1924.84 & 1359.93 & 1.415 \\
          458.sjeng      & 1737.12 & 1662.04 & 1.045 & 1711.17 & 1702.92 & 1.005 \\
          462.libquantum & 2197.98 & 2169.19 & 1.013 & 2234.69 & 2234.48 & 1.000 \\
          464.h264ref    & 2577.43 & 2557.88 & 1.008 & 2509.29 & 2537.91 & 0.989 \\
          471.omnetpp    &  830.17 &  819.39 & 1.013 &  815.68 &  801.21 & 1.018 \\
          473.astar      & 1029.62 & 1005.49 & 1.024 & 1041.58 & 1045.25 & 0.997 \\
          483.xalancbmk  &  798.15 &  823.26 & 0.969 &  764.04 &  767.93 & 0.995 \\\hline
          geomean        & & & 1.056 & & & 1.035 \\\hline\\
      \end{tabular}
      }
      \caption{SPEC06 ref. (25+ trillion instructions) runtime comparison between the baseline core and the Bitmanip-enabled core, as well as with the Zicond-enabled core.}
      \label{tab:spec-bitmanip-zicond}
  \end{table}

We have evaluated these extensions on different benchmarks, all of them booting Linux and running in FireSim\cite{Karandikar18}.
\begin{enumerate}
    \item Bitmanip: We evaluate Bitmanip's implementation on the entire SPEC06 reference suite~\cite{SPEC06}, executing over 25 trillion instructions. Compiling SPEC with gcc 13.2 we obtained an overall speedup of 5.6\% over the baseline, as seen in Table~\ref{tab:spec-bitmanip-zicond}. 
    \item Crypto: Crypto is evaluated using OpenSSL Speed~\cite{OSSL}, a benchmark designed to test the performance of different cryptographic algorithms' performance. The metric used in these benchmarks is throughput---given an input size, it measures the amount of times it can execute a given algorithm. Figure \ref{fig:openssl} shows the speedup over baseline for different input sizes (x-axis) for different OpenSSL algorithms. The reason for us not to use SPEC06 reference~\cite{SPEC06} is that it contains no Crypto instructions.
    \item Zicond is evaluated against SPEC06 ref.\cite{SPEC06} too, compiled with gcc 16.1 provided that previous versions do not emit Zicond instructions. Table \ref{tab:spec-bitmanip-zicond} shows the per-benchmark breakdown improvement and an overall geomean speedup of 3.5\%.
\end{enumerate}

Finally, we have also collected the cost in dollars and tokens of running the loop for the three implemented specifications. The results can be found in the Table~\ref{tab:vext-llm-cost}.

\begin{table}[t]
\centering
\resizebox{\columnwidth}{!}{
    \begin{tabular}{lrrrrrr}
    \hline
    Case & LLM Iters & Input Tokens & Cache Creation. & Cache Read. & Output & Cost (\$)\\
    \thickhline
    Bitmanip & 2 & 136 &   956{,}922 & 22{,}448{,}288 & 110{,}887 & 23.57 \\
    Crypto   & 6 & 205 & 2{,}044{,}094 & 37{,}802{,}171 & 194{,}255 & 46.24 \\
    Zicond   & 1 &  80 &   167{,}106 &  5{,}329{,}628 &  30{,}774 & 5.11 \\
    \hline\\
    \end{tabular}
    }
    \caption{LLM (Claude Opus 4.7) usage and cost of the CHIA loop implementing Bitmanip, Crypto, and Zicond.}
    \label{tab:vext-llm-cost}
  \end{table}

  \textbf{The Power of CHIA:} CHIA enables us to leverage LLMs for specific tasks (processing the specification, making an implementation or debugging) without losing any control in phases where we cannot tolerate the LLM making mistakes (performing verification and performance evaluation, or collecting synthesis results). CHIA also allows each stage of the workflow to be designed using the most suited approach. An example of this can be found in our 3-stage division (see Figure ~\ref{fig:vextdiagram}) where each stage has been constructed with different goals and principles. Still, they all integrate seamlessly in a single functional workflow. Another aspect of CHIA that has made this loop feasible is its ability to harness compute resources from across different physical clusters. In this specific case we have used a mix of on-premises and cloud machines. Verification is computationally intensive, as seen in Figure ~\ref{fig:ext_timelines}, so we have divided it across clusters, while other tasks depend on specific licenses that only exist on-premises. CHIA handles the complexity of making these clusters appear as a single pool of logical resources that the user can employ to run the nodes of any loop.

We provide a getting started guide to deploy the RISC-V ISA extension loop on our documentation website.\footnote{\url{http://chialoops.ai/risc-v-ext-to-rtl}}


\subsection{IPC-aware automated critical path optimization in an out-of-order superscalar RISC-V core}\label{sec:eval:topt}

A processor's maximum operating frequency, and thus its maximum performance, is most directly limited by the longest latency (critical) path which must be traversed by a signal within a single clock cycle. For this reason, optimizing a processor's RTL to minimize the length of this critical path is an important engineering problem. However, breaking down a single long chain of logic without affecting the functionality of the processor can take days of focused human effort, requiring a complex understanding of the signals involved and how they interact with the processor at large. 

\begin{figure}[t]
    \centering
    \resizebox{\columnwidth}{!}{\includegraphics[width=1\columnwidth]{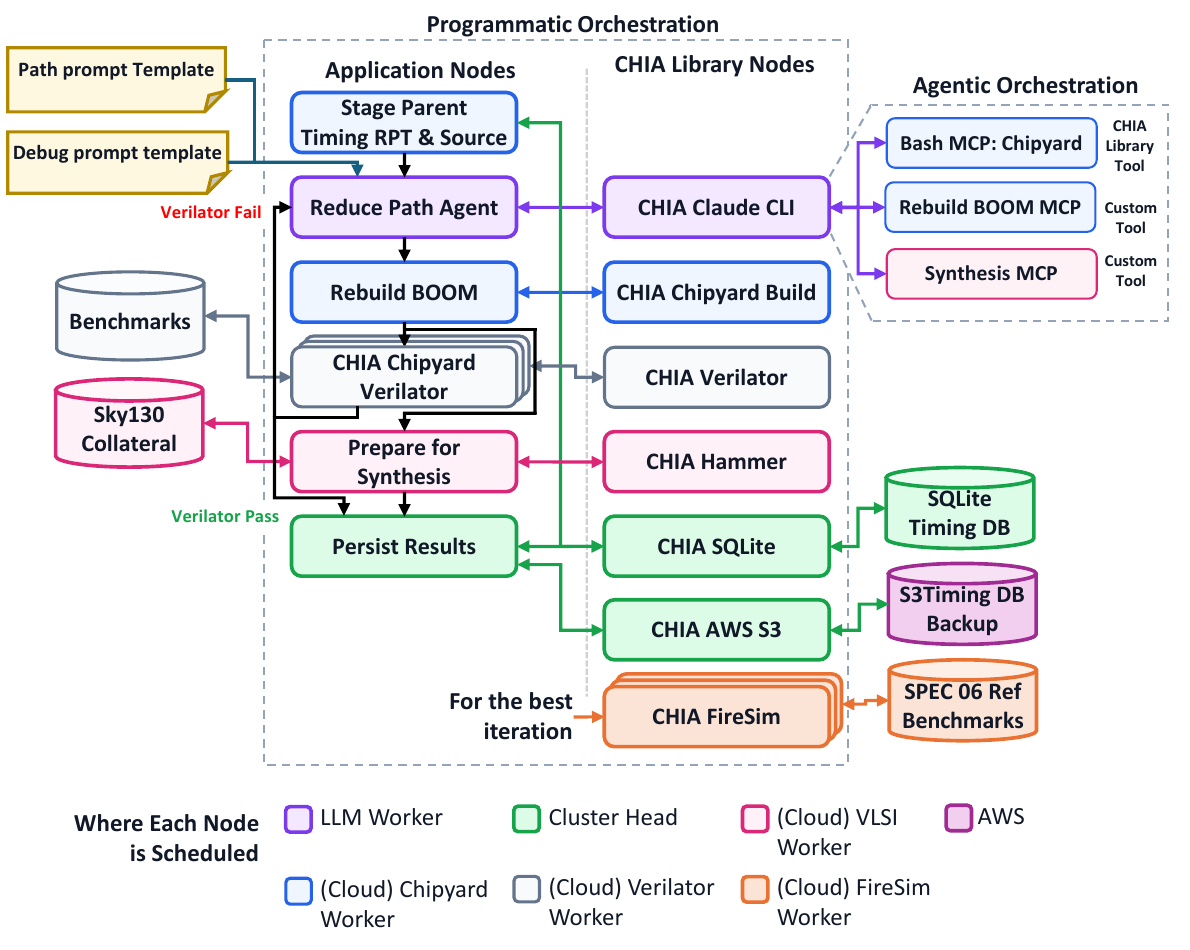}}
    \caption{Architecture of the BOOM critical path optimization CHIA loop. In each iteration, an LLM is asked to reduce the critical path latency of the BOOM core based on timing reports from running gate level Synthesis with the Sky130 PDK~\cite{SKY130}.}
    \label{fig:timingflow}
\end{figure}

\begin{figure*}
    \centering
    \resizebox{\textwidth}{!}
    {\includegraphics[width=1\columnwidth]{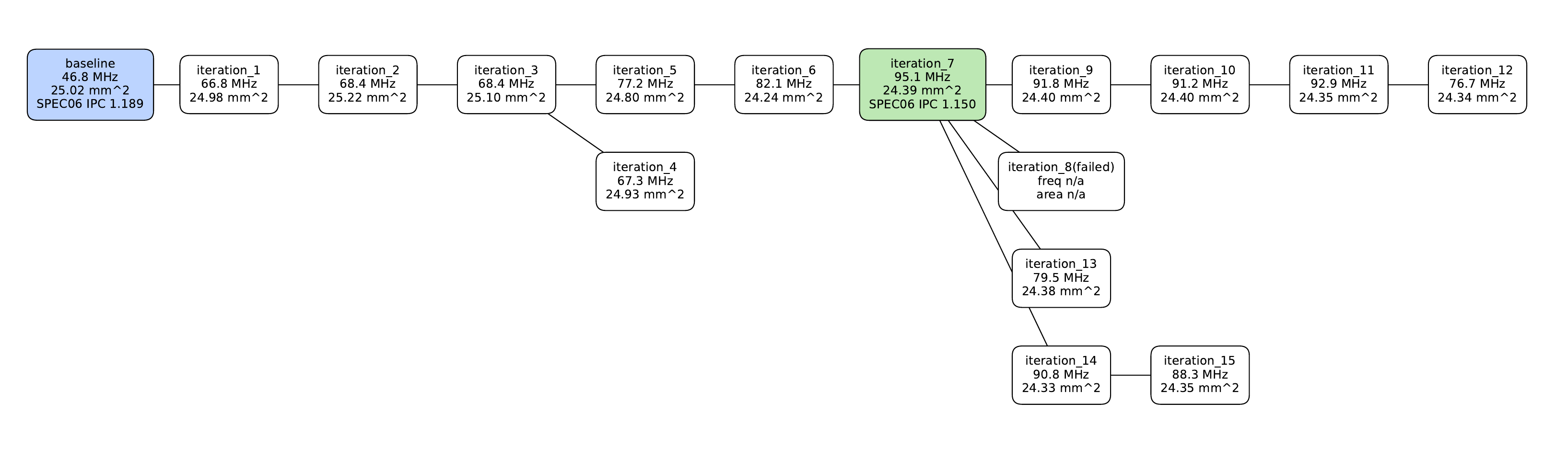}}
    \caption{Timing optimization tree of results, showing maximum achievable frequency and area of each iteration when synthesized in Sky130~\cite{SKY130}. Our best iteration yielded a 2.03x improvement in frequency, with an IPC loss of only 3\% compared to the baseline, for a net Iron-Law speedup of 1.97x. Silicon area changes by a negligible amount.}
    \label{fig:timoptresults}
\end{figure*}

We developed a CHIA loop to take on this task for the 4-wide MegaBOOM core~\cite{Zhao20,Celio18} in Chipyard~\cite{Amid20}. Our loop is described in Figure~\ref{fig:timingflow}. The loop takes a timing report from running a parent iteration's final BOOM core through a commercial gate-level synthesis tool on the open-source Skywater 130nm PDK~\cite{SKY130}, and provides it to the Claude Code CHIA library node using Opus 4.6 as the underlying model. For the first four iterations we asked the agent to reduce critical path with no impact on IPC, but after results plateaued, we changed the focus to iron-law performance (asking the agent to reduce the critical path with as small an IPC reduction as possible). This meant writing a new prompt and turning on IPC collection in the CHIA Verilator nodes, only a few lines of code in CHIA. The LLM in our loop is provided a bash shell in the Chipyard codebase for exploring and editing BOOM, a tool to confirm that BOOM rebuilds correctly, and a tool to run synthesis on small modules within BOOM as a way to test its ideas. Once the LLM is satisfied with its changes, the loop (without AI involved) rebuilds BOOM, and runs it through a large suite of benchmarks in Verilator to verify functionality is unchanged, and to estimate impact on IPC. In parallel, we synthesize the modified BOOM core for area and timing estimates. At the end of each iteration, we persist our results in a SQLite database.

\begin{table}[]
    \centering
    \resizebox{1\columnwidth}{!}{
    \begin{tabular}{lrrrrr}
    \hline
    & \multicolumn{1}{c}{Input Tokens}
    & \multicolumn{1}{c}{Cache Creation}
    & \multicolumn{1}{c}{Cache Read}
    & \multicolumn{1}{c}{Output}
    & \multicolumn{1}{c}{Cost (\$)}
    \\\thickhline
        Average & 106 & 177K & 4.69M & 53.4K & 14.43 \\
        Total (n=14) & 1,478 & 2.48M & 65.7M & 747KM & 202.03 \\\hline\\
    \end{tabular}
    }
    \caption{Average and total LLM token usage and cost for our MegaBOOM timing optimization loop. We achieved an iron-law speedup near 2x for \$202 in API costs. Failed iteration 8 is excluded, though its costs were on the same order of magnitude as the other iterations. The vast majority of tokens input to the model are consumed as "Cache Creation Tokens" and "Cache Read Tokens", and not as "Input Tokens"}
    \label{tab:timoptstats}
\end{table}

We ran our loop until it reached a frequency plateau (less than 5 active days, at less than 8 hours per iteration on average), and during that time it more than doubled the frequency at which we could run the MegaBOOM core in Sky130~\cite{SKY130}, from around 47MHz to around 95MHz in our best iteration (iteration 7, see Figure~\ref{fig:timoptresults}).

For our final measurement of how much IPC was lost, we compare the IPC of iteration 7 with our baseline on all 25 trillion instructions of the SPECint 2006 reference benchmark suite using FireSim~\cite{Karandikar18} (not shown in Figure~\ref{fig:timingflow}). 

The IPC reduction on the SPEC06 ref. suite between the baseline and iteration 7 was 3.28\%, which is very similar to the IPC reduction we see while running the flow with smaller benchmarks in Verilator (3.20\%). \textbf{This yields a net Iron-Law speedup of 1.97x, nearly a doubling in overall performance.} Additionally, we use the SPEC06 runs (over 25 trillion instructions) as validation that the optimizations implemented by the agent did not degrade the functionality of the core. Table~\ref{tab:timoptstats} shows the token usage and cost for running this loop per iteration and in total.

After iteration 7, we see that results generally were flat or regressed. Analyzing the synthesis reports, we find that this plateauing occurs because at this frequency the core is somewhat balanced, with many distinct paths with the same latency, with endpoints spread across the frontend, the load-store unit, and the data cache. On the other hand, we find that our two large regressions share a common critical path, which is unrelated to the agent's changes, and likely emerged because certain synthesis heuristics, for example SRAM macro placement, gave significantly different outputs due to small changes in the inputs.

Invisibly to the agent, we also ran each of our designs through synthesis in a commercial 16nm process. The first of the agent's timing optimizations (restructuring linear-depth structures in the issue queues to log-depth) improved the critical path in the commercial process by about 70\%. However, that was the only improvement in the commercial process from our loop's changes. Put another way, our Sky130 results do not generalize to the advanced commercial node beyond the first optimization. As the 16nm process is proprietary, we could not use it in this case study, since information from the PDK cannot be given to public frontier models. In section~\ref{sec:fut}, we will discuss this problem in more detail, and its implications for assessing QoR in AI-guided hardware design.

\textbf{The Power of CHIA: }This case study heavily leverages the reusability of CHIA's library nodes, and the ability to carve out pieces of one workflow and incorporate them into another. Many of the nodes used in this case study, including building BOOM's Chisel, simulation in Verilator and FireSim, and gate-level Synthesis, rely on CHIA library nodes. Furthermore, this case study was able to share many of its application nodes with the previous case study (implementing new RISC-V ISA extensions in the BOOM core) because the two are operating on the same core, meaning much of the pre- and post-processing required for each of the library nodes is the same, despite the goals of the two case studies being very different.

This case study was also able to heavily leverage CHIA's support for public cloud services. In particular, we used cloud machines to run Chipyard Build, Verilator, Hammer/Synthesis, and FireSim, and we also used CHIA's AWS S3~\cite{AWSS3} support to keep our database backed up in the cloud.


We provide a detailed walkthrough of the IPC-aware critical path optimization case study on our documentation website.\footnote{\url{http://chialoops.ai/critical-path-opt}}

\subsection{Agentic Architectural Discovery with Evolutionary Coding Agents}
\definecolor{codegreen}{rgb}{0.0,0.45,0.2}

\begin{figure*}[t]
    \centering
    \includegraphics[width=0.65\textwidth]{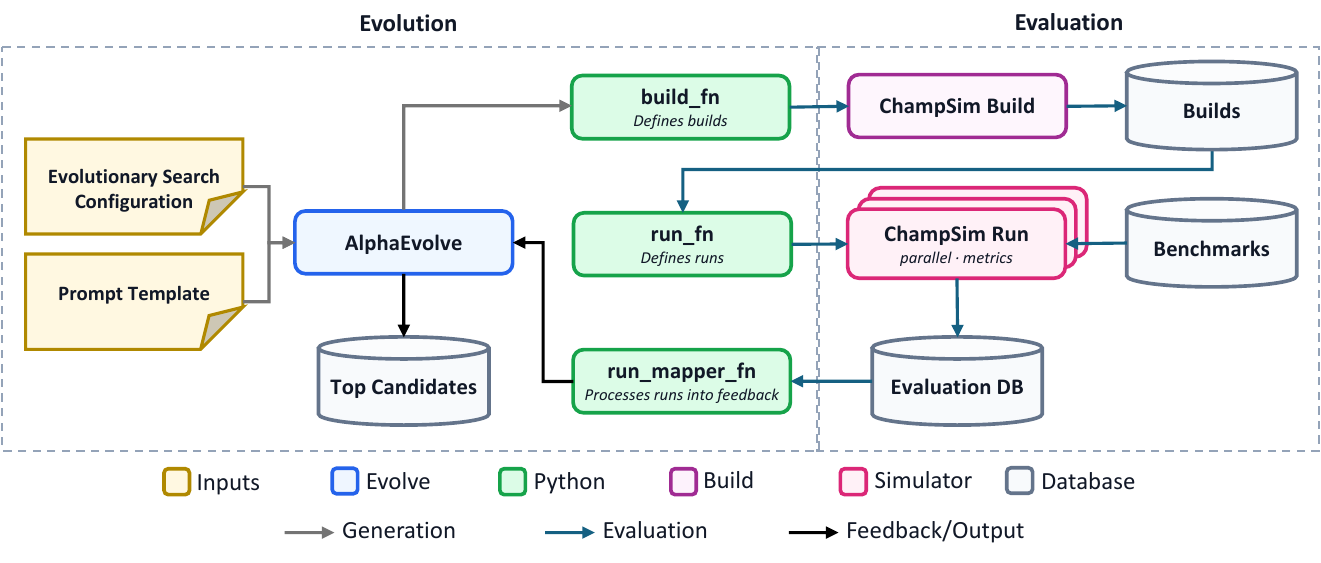}
    \caption{Architecture of an illustrative agentic discovery CHIA loop mirroring the simpler flows shown in prior work.}
    \label{fig:simple-discovery-flow}
\end{figure*}

\begin{figure*}[htbp]
    \centering
    \includegraphics[width=\textwidth]{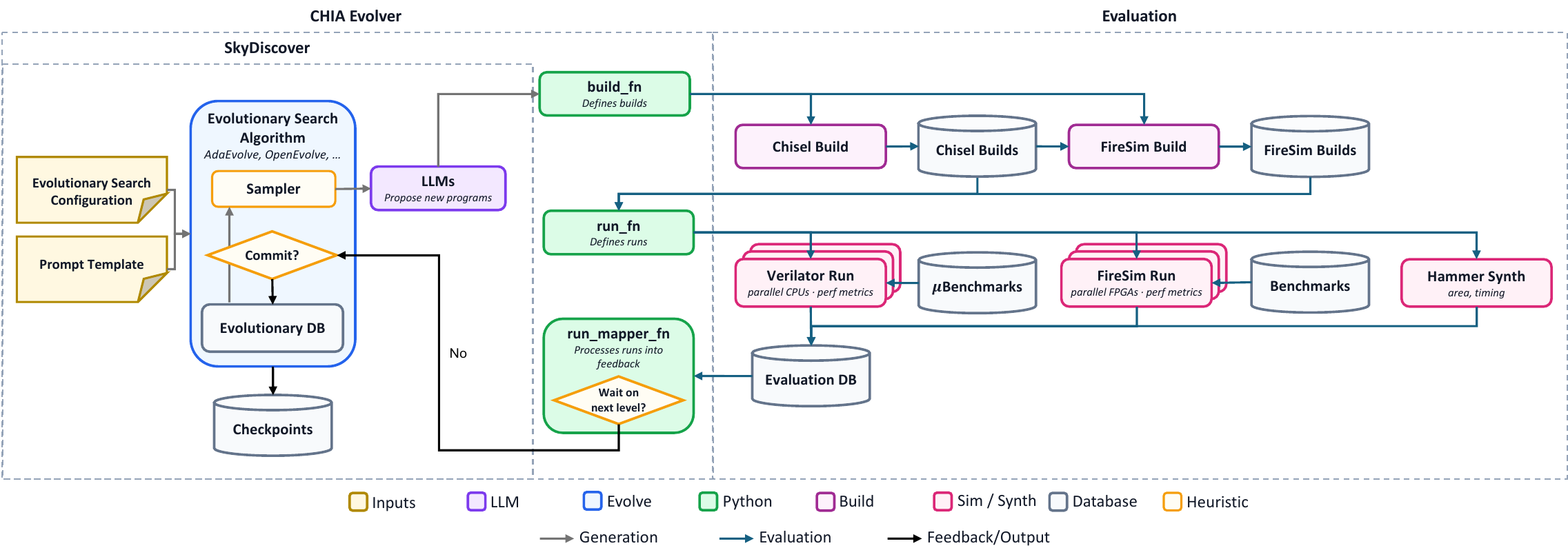}
    \caption{Architecture of an RTL discovery CHIA loop featuring the ability to choose evolutionary coding agents along with a cascade of evaluators including Verilator, FireSim, and commerical synthesis tools via Hammer. In practice, we would expect to add CHIA LLM nodes throughout the flow to act upon the comprehensive data generated by the cascade of evaluators.}
    \label{fig:complex-discovery-flow}
\end{figure*}

AI techniques such as evolutionary coding agents have shown significant promise in the discovery of novel ideas and approaches across scientific domains including the computing stack ~\cite{cheng2025barbarians, novikov2025alphaevolve, georgiev2025mathematical}.

Within computer architecture, ArchAgent~\cite{anonymous2025archagent, Gupta26} combined AlphaEvolve with a highly distributed simulation cluster to discover novel cache replacement policies. These generated policies outperformed prior human-designed state-of-the-art policies in a competition setting while far outpacing their estimated development times. 


\subsubsection{Challenges in Using Agentic Discovery Flows}

While we have observed success in applying agentic discovery flows to problems in computer architecture and computer systems, significant challenges still remain. These challenges can be divided into three broad categories: 
\begin{enumerate}
    \item \textbf{Low Fidelity and Low Performance Evaluations}: AI agents are prone to reward hacking. In an early experiment in ArchAgent, AlphaEvolve was observed to reward hack an improperly enforced assertion to drop memory traffic and claim performance improvements. 
    
    More broadly, evolutionary coding agents are sensitive to evaluation fidelity - the quality and accuracy of feedback significantly affects the utility of generated designs. For instance, microarchitectural simulators such as ChampSim are unable to provide reliable feedback on power, area, and critical paths. This brings into question the hardware realizability of designs generated with feedback from such models. Substantial human effort is required to reason about and verify large volumes of such AI-generated designs. This defeats the purpose of deploying an automated flow in the first place. 
    
    For any given time frame, an evolutionary coding agent's ability to explore a design space is fundamentally limited by the latency of evaluating proposed candidates. Thus, feedback from fast and high-fidelity simulators is essential for microarchitecture discovery. However, ChampSim required days to simulate running a single billion-instruction length SimPoint of a SPEC06 workload on a quad-core processor ~\cite{Gupta26}. One could easily compile a larger design onto an FPGA-accelerated RTL simulator such as FireSim (as shown in our other case studies) and execute much longer workloads in lesser time while simultaneously gaining much higher fidelity feedback.
    \item \textbf{Unergonomic Flow Development}: Researchers report that designing and delivering a successful discovery flow is extremely challenging, especially in the presence of long-running evaluations (often observed in hardware/software co-design research) that can cause errors to materialize at timescales of hours to days. Machine errors and downtime also become an increasingly relevant concern, and flow interruptions can discard hours to days of valuable intermediate results. Similarly, it can take hours to days to observe the effects of a single change. Experimenting with a sequence of such individual changes on flows with fixed configuration parameters or heuristics (such as those for population sampling or commit) costs extensive amounts of time and effort. 
    \item \textbf{Limited Flow Exploration}: Relatedly, existing microarchitecture discovery flows as seen in ~\cite{Gupta26, blasberg2026agenticarchitectagenticai} offer point solutions in the design space of agentic approaches to microarchitecture discovery. The broader composition of agents for problems in computer architecture (as shown in our other case studies) remains underexplored in the context of evolutionary discovery. For instance, current evolutionary discovery flows have an evolutionary coding agent as the topmost flow node. However, this is not necessary and an evolutionary coding agent could instead be more effective as a piece in a larger agentic flow. We believe such hypotheses remain untested in part due to the churn and tedium experienced in developing long-running discovery flows. 
\end{enumerate}

Rather than showing a specific result in this section, we demonstrate how existing flows like ArchAgent can be expressed in CHIA, how CHIA enables addressing many of the challenges faced in conventional evolutionary coding agent flows, and how CHIA facilitates further research towards groundbreaking discovery flows.

\subsubsection{"Classic" evolutionary discovery flows in CHIA}
First, we look at the design of an illustrative CHIA loop implementing an evolutionary flow similar to the one seen in ArchAgent. In order to do this, we provide ChampSim nodes and an AlphaEvolve node in CHIA, organizing them as shown in Figure~\ref{fig:simple-discovery-flow}.  

The user selects configurations for the ChampSim nodes, configuring the simulated SoC, and the AlphaEvolve node, configuring various evolutionary search parameters. The user also provides a few lightweight Python functions. The function \lstinline[language=python, basicstyle=\ttfamily\color{codegreen}]{build_fn} is used to define how binaries are built on a ChampSim Build node and built binaries are passed further into evaluation. 
In the case of evolution feedback, for example, ChampSim Run nodes could be set up to execute a selection of workload traces from a workloads database executing for a short number of warmup and detailed simulation instructions. This configuration is defined in the function \lstinline[language=python, basicstyle=\ttfamily\color{codegreen}]{run_fn}. 
Metrics such as instructions-per-cycle, misses-per-kilo-instruction, prefetcher coverage/accuracy can be aggregated and provided as feedback using the function \lstinline[language=python, basicstyle=\ttfamily\color{codegreen}]{run_mapper_fn}.

\subsubsection{Advanced high-fidelity evolutionary discovery flows in CHIA}
CHIA facilitates designing much more powerful agentic discovery flows by providing extensible interfaces and a rich library of tool nodes that enable incorporating comprehensive and actionable feedback into the evolutionary search:
\begin{itemize}        
    \item \textbf{Evaluator Nodes}: CHIA enables the composition of diverse evaluation backends such as functional models (Spike), microarchitectural simulators (gem5, ChampSim), RTL simulators (Verilator) and emulators (FireSim), and physical design tools (via Hammer). These tools can be used to provide high-fidelity performance (frequency and cycle-by-cycle), power, and area metrics to evolutionary coding agents.
    \item \textbf{Evolver Node}: SkyDiscover~\cite{Liu26} is a flexible framework for AI-driven scientific and algorithmic discovery with support for widely-used large language models, the Claude code coding agents, and a variety of evolutionary coding agents such as AdaEvolve~\cite{cemri2026adaevolveadaptivellmdriven}, EvoX~\cite{liu2026evoxmetaevolutionautomateddiscovery}, GEPA~\cite{agrawal2026gepareflectivepromptevolution}, and OpenEvolve~\cite{openevolve}. CHIA ships with an Evolver node providing flexible interfaces around SkyDiscover and which allow it to interact with the rest of the framework. We provide support for AlphaEvolve~\cite{novikov2025alphaevolve} via the AlphaEvolve Google Cloud API \cite{nawalgaria2026alphaevolve} which is added as an external backend to SkyDiscover inside the CHIA Evolver node (referred to as AlphaEvolve node above).
\end{itemize}

Figure~\ref{fig:complex-discovery-flow} shows an example of a more data-rich and flexible RTL discovery flow that lets the user pick between various evolutionary coding agents and features a cascade of evaluators providing Verilator, FireSim, and Hammer ASIC Quality-of-Results data for feedback, all  productively composed with CHIA. 

Traditionally, building and debugging a flow like this would be much more difficult than constructing one using a single kind of evaluator that executes purely in software (such as a microarchitectural simulator). CHIA not only provides a library of battle-tested tools, it also makes much-needed improvements to the development flow. In an early experiment with AdaEvolve, we made a typographical error in configuring its guide model. This error disabled paradigm generation, one of AdaEvolve's most valuable features that counters flow stagnation. Paradigm generation errors were first observed 12 iterations and 4 hours into the experiment and significantly slowed down its rate of progress. In the absence of CHIA, our options were to restart from scratch or wait for the next checkpoint at 25 iterations. In such a situation, CHIA's cache and bypass feature allows us to pause the flow, fix the error, fast-forward to the previous error point using cached CHIA function outputs, and resume normal execution. CHIA can also flexibly manage its clusters such that we would not have to bring down or restart our entire cluster to correct this error. 
 
Similarly, CHIA allows further composition of evolutionary flows into broader agentic flows (similar to the ones described in previous case studies), targeting discovery across the computing stack with the ability to span multiple levels of representation such as evolving software programs, microarchitecture models, RTL implementations, and circuit designs. In combination with improvements to flow development ergonomics, we believe CHIA enables researchers to push the frontier of agentic discovery flow design patterns and explore open questions in the design and use of such flows for hardware-software co-design. 

\textbf{The Power of CHIA}: This case study showcases CHIA's ability to compose powerful discovery flows from reusable parts: its extensible node library lets us assemble cascades of high-fidelity evaluators—from functional models and microarchitectural simulators to RTL simulators, emulators, and physical design tools—and freely swap between evolutionary coding agents, turning what were once bespoke, monolithic flows into building blocks for discovery across the computing stack. Just as importantly, CHIA relieves the churn and tedium of developing long-running discovery flows, with features like cache-and-bypass and flexible cluster management that help researchers recover from the errors and interruptions that inevitably surface over hours- to days-long evaluations. Together, these capabilities let researchers move past today's point solutions and explore the broader, largely untested design space of agentic discovery flows.


We provide a detailed walkthrough of the architectural discovery with evolutionary coding agents case study on our documentation website. \footnote{\url{http://chialoops.ai/discovery-basic}}\textsuperscript{,}\footnote{\url{http://chialoops.ai/discovery-advanced}}\label{sec:eval:disc}

\subsection{Automatically addressing GitHub issues in the CIRCT compiler in a project maintainer friendly way}

\begin{figure}
    \centering
    \resizebox{\columnwidth}{!}{\includegraphics[width=1\columnwidth]{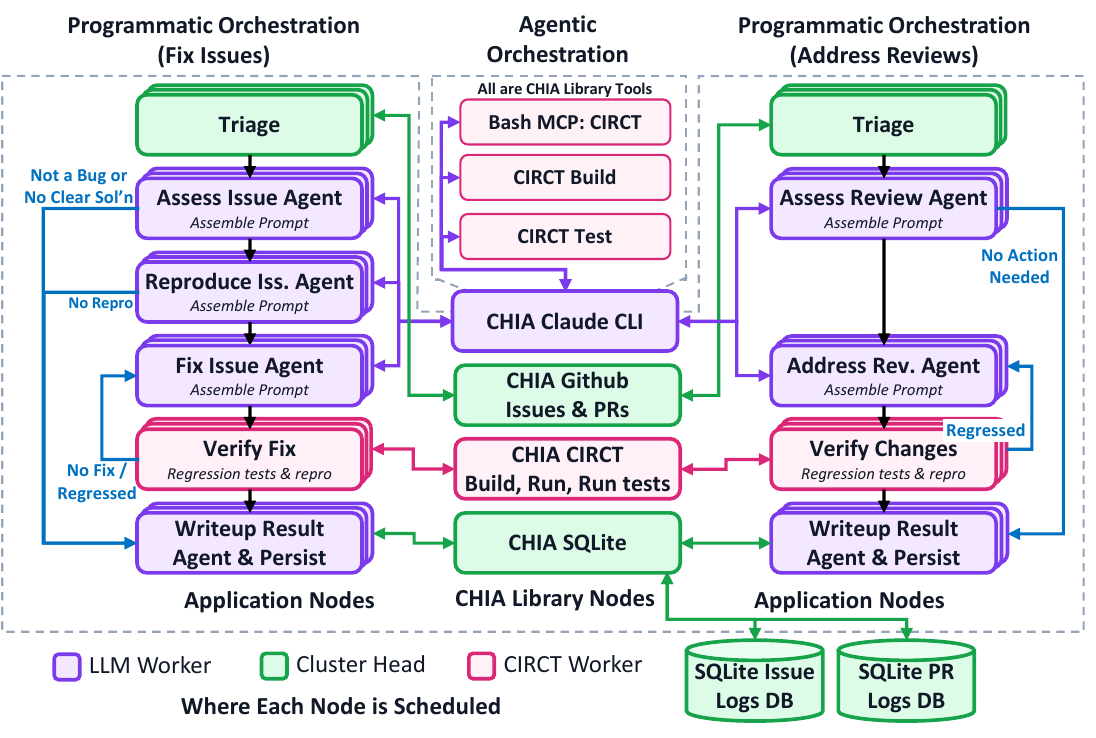}}
    \caption{Architecture of the CIRCT GitHub issue fixing and PR response CHIA loops. Each loop uses a sequence of prompts to an agent to assess an issue or comment and take the correct action in a principled way.}
    \label{fig:circtflow}
\end{figure}

AI tools have enabled an increase in outside contributions to large open source GitHub repositories, including new issue reports and pull requests. As noted in the LLVM project's AI tool use policy, when these contributions are done entirely by AI in an unprincipled way and without human review, this "extracts work from [maintainers] in the form of design and code review"~\cite{LLVMAI}. While CHIA cannot force a human contributor to review the work of their agent, it can be used to force the agent to develop contributions in a more principled way. To demonstrate this, we built a CHIA loop, as seen in Figure~\ref{fig:circtflow}, to solve issues for LLVM's CIRCT project, which compiles high-level hardware description languages into SystemVerilog and RTL simulators. Notably, CIRCT is used in Chipyard for lowering its Chisel-based generators down to System Verilog. To reduce maintainer burden rather than increase it, we focus our efforts only on issues that demonstrate bugs, and where the conversation surrounding the issue has left no ambiguity about the correct resolution for the bug.

The loop works as follows. First, a GitHub issue is triaged heuristically, filtering out issues with no code blocks, and those with an open PR referencing the issue (which implies a solution to the issue has already been proposed). Additionally, since we limit our scope to bugs and not feature requests, we avoid any issues labeled as feature requests. An issue is then assessed by an LLM to determine whether (1) the issue is actually a bug and (2) there is a single clear solution to the issue. We do this to make sure that any changes align with our narrow scope of bugs with a single, clear resolution. For issues where this is the case, an agent then tries to reproduce the bug (outputting a bash file which, when run, reproduces the bug) and if successful, another agent tries to fix the bug. These agents are given tools to modify and build CIRCT source, and run tests using CIRCT binaries to validate their changes. After this, we take control back from the agents, and without any AI involvement, the loop verifies the fix by running CIRCT's full suite of regression tests, and the reproduction bash file, to ensure the bug is fixed.

While we do have the LLM write up a PR for the issue, to minimize extractive work and to comply with LLVM's AI policy, we had a human review each change in detail to ensure quality and hand-write pull requests to submit to the upstream CIRCT repository. It took a graduate student with novice-level familiarity with CIRCT an hour per issue to review the agents' changes and ensure they were high enough quality to submit pull requests. It took about 10 more minutes per issue to prepare the pull requests. Additionally, in compliance with CIRCT's AI usage policy, we did not submit pull requests for any issues marked as "good first issues". We also built a loop for addressing reviewer comments, which follows a similar pattern, first assessing comments for action items, and taking action where necessary, before validating in an environment with no agents. We used this loop to privately address comments we expected to receive, and to address actual comments from maintainers.

\begin{table}[]
    \centering
    \resizebox{1\columnwidth}{!}{%
    \begin{tabular}{ccccc}
       \hline Issue No. & Assessed As & Reproducible & Fixed & Pull Request Submitted \\\thickhline
       \#2266 & Not a bug & - & - & - \\
       \#2669 & Not a bug & - & - & - \\
       \#4354 & Bug & Yes & Yes & No (Good First Issue) \\
       \#4396 & Not a bug & - & - & - \\
       \#4649 & Fix unclear & - & - & - \\
       \#5626 & Fix unclear & - & - & - \\
       \#5789 & Bug & Already fixed & - & - \\
       \#6226 & Bug & Already fixed & - & - \\
       \#6740 & Bug & Yes & Yes & No (Good First Issue) \\
       \#7127 & Not a bug & - & - & - \\
       \#7388 & Bug & Yes & Yes & Yes (Merged) \\
       \#7531 & Fix unclear & - & - & - \\
       \#7949 & Bug & Yes & Yes & Yes (Merged) \\
       \#8508 & Fix unclear & - & - & - \\
       \#10104 & Bug & Yes & Yes & Yes (Merged) \\
       \#10571 & Not a bug & - & - & - \\\hline\\     
    \end{tabular}
    }
    \caption{Characterization of the 16 randomly selected CIRCT GitHub issues at which we targeted our loop. Our workflow correctly categorized nine issues as not bugs or bugs with unclear solutions and two issues as already fixed, and successfully fixed the remaining five issues.}
    \label{tab:circtissuesresults}
\end{table}

\begin{table}[]
    \centering
    \resizebox{1\columnwidth}{!}{
    \begin{tabular}{lrrrr}
    \\\hline
    \multicolumn{1}{c}{Metric} &
    \multicolumn{1}{c}{Assess} &
    \multicolumn{1}{c}{Reproduce} &
    \multicolumn{1}{c}{Fix} &
    \multicolumn{1}{c}{Writeup}
    \\\thickhline
        Input Tokens & 7 & 27 & 68 & 2 \\
        Cache Creation Tokens & 10.5K & 21.5K & 74.6K & 9.4K \\
        Cache Read Tokens & 95.8K & 648K & 3.58M & 8.1K \\
        Output Tokens & 3.2K & 8.8K & 30.0K & 1.2K \\
        Cost (\$) & 0.40 & 0.68 & 3.01 & 0.09 \\\hline\\
    \end{tabular}
    }
    \caption{Average LLM token usage and cost for each agent stage in our issue fixing loop. The vast majority of tokens input to the model are consumed as "Cache Creation Tokens" and "Cache Read Tokens", and not as "Input Tokens".}
    \label{tab:circtstats}
\end{table}

To minimize extractive work, we discussed closely with maintainers of the project to ensure our contributions were of a reasonable quantity and quality. In the end, we ran the loop on 16 randomly selected GitHub issues, and its results are outlined in Table~\ref{tab:circtissuesresults}. Of the 16 issues the loop examined, 11 were determined to be bugs, four of which did not have unambiguous solutions, leaving seven issues entering the reproduction phase. In this stage, two of the seven could not be reproduced, indicating that they had already been fixed in upstream CIRCT. Our loop successfully fixed the remaining five bugs, and we submitted pull requests for the three of them that were not labeled as good first issues. Table~\ref{tab:circtstats} shows the token consumption and cost of each phase of our loop for solving CIRCT issues. Our loop considered all 16 issues in parallel, leveraging CHIA's scalability, and finished in less than 45 minutes.

Each of these three PRs received feedback from reviewers, and we used our loop for addressing reviewer comments to handle their requested changes. All three of our PRs have been merged into the main branch of the CIRCT repository.

We agree with the loop's conclusions for each of the 16 issues. However, as we were prototyping the loop, we came across a situation where our assessor agent determined that issue \#8508 had a clear solution, and our fixer agent successfully fixed the issue, but the assessor and the fixer disagreed about the correct solution. To remedy this, in our final loop, the assessor was required to enumerate all good solutions for a given issue and, if more than one could be described, declare that issue did not have a clear and unambiguous solution. After making this change, the assessor agent reliably declared issue \#8508 had no clear solution.

\textbf{The Power of CHIA: } We see the power of CHIA in a few major ways for this workflow. Firstly, we were able to prototype our flow at a very small scale, with only a single agent and CIRCT instances. Then, only changing 3 lines of code across the cluster configuration and Python parameters, we were instantly able to run as many issues in parallel as we wanted. We also see, in this case study, CHIA's ability to balance agentic orchestration with programmatic orchestration. In each agentic step, the agent is given runway to orchestrate complicated sequences of tasks in order to make decisions about the issue it is fixing. But, to ensure fixes are principled, programmatic orchestration is used to connect the narrowly-scoped agents together, and more importantly, to verify that the changes work as well as the agents claim.

We provide a detailed walkthrough of the CIRCT issue-fixing case study on our documentation website. \footnote{\url{http://chialoops.ai/circt-pr-fixer}}\label{sec:eval:circt}

\section{Related Work}\label{sec:relwork}

Applying artificial intelligence in the context of hardware design is a booming research area.
In particular, many papers have explored using agents to generate RTL~\cite{VerkorTeam26, theverkorteam26_2, Deng26, Wu26, Zhang25, zhao24, zhao25, wang25_2, Liu24, Gai25}. Strategies for this vary greatly, from e-graph rewriting~\cite{Zhang25} to multi-level summarization~\cite{zhao25}. Each of these works uses a bespoke design loop, making their work difficult to compare and impossible to interoperate. Performing these experiments in CHIA would be easier, allow for fairer comparison, and streamline merging successful ideas together.

Discovery of novel architectural ideas is also a subject of growing interest in the architecture community~\cite{anonymous2025archagent, Gupta26, Blasberg26, Sankaralingam26}. Some recent works used evolutionary algorithms to discover novel microarchitectural techniques to optimize performance in out-of-order superscalar cores~\cite{anonymous2025archagent, Gupta26, Blasberg26}. Similarly, one work explored GPU architecture design space exploration by composing DSE rules with LLMs ~\cite{Zhang26}. One work sought to develop a unified sandbox for connecting search algorithms to architectural simulators, in order to streamline ML-aided design space exploration~\cite{Krishnan23}. Another work uses an LLM to guide multi-objective black-box optimization towards the Pareto-frontier, significantly reducing time wasted exploring infeasible points~\cite{pandit25}.

Multiple papers have demonstrated the strong capabilities of LLMs to generate high-performance kernels for GPUs and accelerators~\cite{Hong25, Xu26}. Likewise, LLMs have been used to provide humans with better understandings of kernel~\cite{Davis26} and CPU cache performance~\cite{Mhapsekar26}. LLM-based oracles like this can easily be built in CHIA.

Building off of the success of benchmark suites evaluating the ability of LLMs to generate software~\cite{Jimenez24, Ouyang25}, and noticing the proliferation of work targeting LLMs for hardware design, many hardware focused benchmark suites have been developed. Some of these suites emphasize evaluating agentic co-design~\cite{Tsai26, Alvanaki25}, RTL generation~\cite{Lu24, Pinckney25, Yu26, Guo25} and high-level synthesis~\cite{AbiKaram25}. Building these benchmark suites on top of CHIA would streamline the process of comparing different design loops, and we are interested in integrating these suites into CHIA. 

Recent work has focused on LLM guided usage of EDA tools~\cite{Chen26, He23, li23, Wang25}. These ideas range from MCP EDA tool servers ~\cite{Wang25} to standardized execution frameworks for EDA~\cite{Chen26}. We also hope to encapsulate tools like this into CHIA.

A growing body of work targets general agentic frameworks. CHIA is built on top of Ray \cite{Moritz18}, and adds customizable profiling as well as fault tolerance and testing features through the bypassing and caching capabilities. CHIA also adds easy setup for hetereogeneous clusters that have multiple different workers per machine as well as clusters split across on-premises and cloud compute. Other distributed agent frameworks lack expressive control flow semantics that encompass cyclic workflows \cite{apache15}, distributed fault tolerance support \cite{microsoft25}, or resource-aware scheduling \cite{microsoft25}. Agents and agent platforms designed for a single machine generally lack support for distributed execution \cite{langgraph, openclaw, Antigravity, claudecode, anomalyco25}.

The CHIA framework is the first of its kind. As we note, CHIA synergizes well with many of these new tools like benchmark suites and EDA frameworks, by providing a substrate for easily developing loops which interoperate these tools with others in complex ways. Furthermore, CHIA provides a simple, efficient, and streamlined substrate for running the experiments performed in much of the work discussed above.

\section{Discussion and Future Work}\label{sec:fut} \label{sec:disc}
While developing and using CHIA, it has become clear that the bottlenecks in hardware design pipelines have shifted. As the process of turning an idea into a preliminary implementation speeds up, evaluating the performance and QoR of that idea as well as verifying the idea takes the bulk of the time. CHIA on its own attacks this problem with parallelism, but Amdahl's law tells us that parallelism alone is not enough. Rapid, high-fidelity evaluation and verification of hardware designs will be key to taking full advantage of the capabilities of agentic AI. This means we need shorter representative benchmarks, faster simulation methods and PPA estimation tools, and rapid verification methods without sacrificing fidelity and reliability.

Our field also faces challenges with privileged information. Since proprietary data usually cannot be fed to public frontier model-based agents, open PDKs~\cite{Stine07, Clark16, SKY130} are used in CHIA loops in this paper to demonstrate the power of having PDK data in-the-loop in an agentic flow. Unsurprisingly, these PDKs do not always serve as predictive guides when the ultimate goal is to produce a chip in an advanced process. Therefore, it is critical to develop methodologies to sufficiently abstract quality-of-result and other relevant data before passing it to the largest frontier models. Alternatively, having capable, small, locally hosted models could solve this problem, since sensitive data remains private. CHIA already supports local model hosting and will continue to improve support for it moving forward. 

Finally, we are actively dogfooding CHIA across a variety of projects and we plan to continue developing and supporting CHIA in the open-source.

\section{Conclusion}\label{sec:conc}

CHIA reframes the central challenge of AI-driven hardware/software co-design: rather than treating each application of AI as a bespoke, one-off experiment, it makes the construction, deployment, and rigorous evaluation of the co-design flow itself a first-class concern. By expressing agentic design flows as CHIA loops and backing them with infrastructure for isolation, profiling, evaluation, and reliable execution across hundreds of heterogeneous systems, CHIA lowers the barrier to conducting principled, reproducible science at scale. The five case studies presented here demonstrate the capabilities of CHIA, spanning evolutionary architectural discovery, LLM-driven RTL implementation, automatic simulator alignment, IPC-aware critical path optimization, and automated GitHub-issue fixing.
Across the case studies, we hold the agents operating within the CHIA loops to a high standard of verification and validation. 
For example, the RTL implemented by agents delivers substantial performance improvements while successfully executing the entire 25+ trillion instructions of SPEC06-Ref in the context of a complete out-of-order superscalar microprocessor in a RISC-V system-on-chip. These agent-written implementations achieve their improvements while meeting or improving frequency and area constraints in open-source and commercial ASIC PDKs.

We emphasize that these case studies are illustrative rather than exhaustive. CHIA is designed as a foundation: our goal is to give the architecture, systems, compiler, and VLSI communities a shared, open-source substrate on which to build, compare, and extend AI-infused co-design workflows. As agentic AI continues to mature, we believe frameworks like CHIA will be essential to translating that progress into durable, verifiable advances in hardware/software co-design, and we hope that CHIA serves as a useful platform for the community upon which to build new AI-driven co-design flows.

\section{Acknowledgments}\label{sec:ack}
We thank Schuyler Eldridge and the CIRCT code reviewers for their help in shepherding the merging of our CIRCT pull requests.

The information, data, or work presented herein was funded in part by the Jean and Hing Wong Foundation Faculty Fellowship, Samsung Advanced Institute of Technology (SAIT), and SLICE Lab industrial sponsors and affiliates. Any opinions, findings, conclusions, or recommendations in this paper are solely those of the authors and do not necessarily reflect the position or the policy of the sponsors. This material is also based in part upon work supported by the National Science Foundation Graduate Research Fellowship Program under Grant No. DGE 2146752. Any opinions, findings, and conclusions or recommendations expressed in this material are those of the author(s) and do not necessarily reflect the views of the National Science Foundation.

\bibliographystyle{ACM-Reference-Format}
\bibliography{chia-bib}

\end{document}